\documentclass[twocolumn]{aastex}
\usepackage{amsmath}
\usepackage{amsbsy}

\newcommand{\nflares}{3137 }
\usepackage{booktabs}

\shorttitle{A Database of Flare Ribbon Properties From SDO I: Reconnection Flux}

\shortauthors{Kazachenko et al.}

\begin{document}
\title{A Database of Flare Ribbon Properties From\\ \emph{Solar Dynamics Observatory} I: Reconnection Flux}
\author{Maria~D.~Kazachenko\altaffilmark{1}, Benjamin~J.~Lynch\altaffilmark{1}, Brian~T.~Welsch\altaffilmark{2},
Xudong~Sun\altaffilmark{3}}
\altaffiltext{1}{Space Sciences Laboratory, University of California--Berkeley, Berkeley, CA 94720, USA}
\altaffiltext{2}{Natural \& Appl. Sci., Univ. of Wisc. - Green Bay}
\altaffiltext{3}{W. W. Hansen Experimental Physics Laboratory, Stanford University, Stanford, CA 94305, USA}
\email{kazachenko@ssl.berkeley.edu}
\begin{abstract}
We present a database of $3137$ solar flare ribbon events corresponding to every flare of GOES class 
C1.0 and greater within 45 degrees from the central meridian, from April 2010 until April 2016, 
observed by the \emph{Solar Dynamics Observatory}. For every event in the database, we compare the GOES peak X-ray flux with corresponding active-region and flare-ribbon properties.
We find that while the peak X-ray flux is not correlated with the active region unsigned magnetic 
flux, it is strongly correlated with the flare ribbon reconnection flux, flare ribbon area, and the fraction of active 
region flux that undergoes reconnection. We find the relationship between the peak X-ray flux 
and the flare ribbon reconnection flux to be $I_\mathrm{X,peak} \propto \Phi_\mathrm{ribbon}^{1.5}$. 
This scaling law is consistent with earlier hydrodynamic simulations of impulsively heated flare loops. 
Using the flare reconnection flux as a proxy for the total released flare energy $E$, we find that the occurrence frequency of flare energies follows a power-law dependence: $dN/dE \propto E^{-1.6}$ for $10^{31}<E<10^{33}$ erg, consistent with earlier studies of solar and stellar flares. The database is available online and can be used for future quantitative studies of flares.
\end{abstract}


%
%
\section{Introduction}

Solar flare emission over a wide range of electromagnetic wavelengths is a result of the rapid conversion of free magnetic energy stored in the sheared and/or twisted 
magnetic fields of active regions (ARs) \citep{Priest1981,Forbes2000,Fletcher2011,Hudson2011,Shibata2011,Kazachenko2012}. 
Large flares are often accompanied by coronal mass ejections \citep[CMEs, ][]{Andrews2003}, 
but not all flares are associated with CMEs \citep{Hudson2011, Sun2015}, and some 
CMEs occur without any flare emission \citep{Robbrecht2009, DHuys2014}. 
The total energy released during solar flares typically ranges between $10^{29}$ to $10^{32}$ ergs \citep[e.g.][]{Emslie2012}. 

Flare ribbons are enhanced H$\alpha$ and 1600\AA{} UV emission intensity 
structures in the transition region and the upper chromosphere at the height of approximately $2000$~km. 
The enhanced emission is thought to occur in response to the precipitation of non-thermal particles accelerated either directly or indirectly by magnetic reconnection \citep{Forbes2000,Fletcher2011,Qiu2012,Graham2015,Longcope2014,Li2014,Li2017,Priest2017}. Therefore, the flare ribbons correspond to the 
footpoints of newly reconnected flux tubes in the flare arcade.

\begin{figure}[!tb]
 	\centerline{ \includegraphics[width=0.5\textwidth]{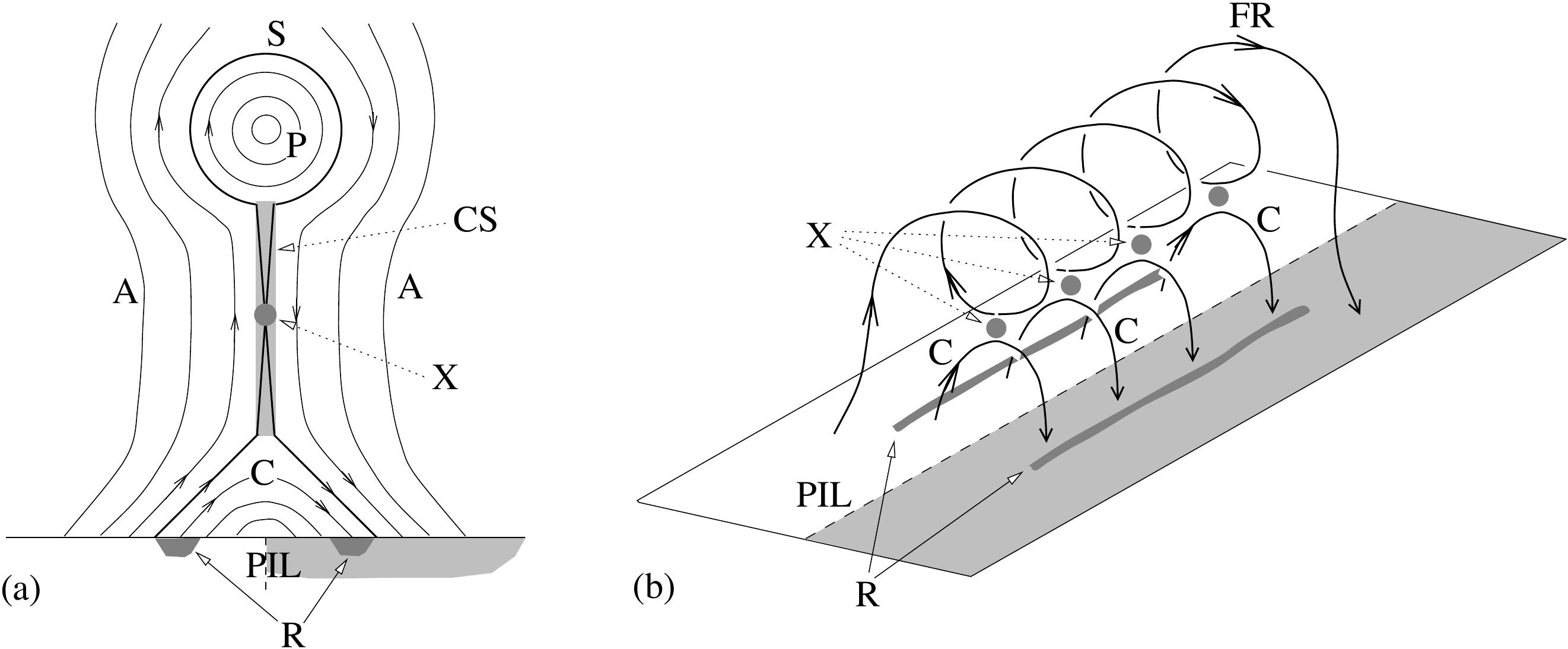} }
 	\caption{Basic elements of the CSHKP two-ribbon flare model in 
 	(a) two dimensions \citep[2D; from][]{Forbes2000} and (b) three 
 	dimensions \citep[3D; from][]{Longcope2007}. Here, `R' 
 	indicates the location of the flare ribbons, `CS' the current sheet, `A' the overlying arcade, `P' the erupting plasmoid, `FR' the 3D flux rope,
 	`PIL' the polarity inversion line, `X' the site(s) of magnetic 
 	reconnection, `S' the separatrix boundary of the erupting CME flux 
 	rope, and `C' the coronal flare loops formed by magnetic 
 	reconnection. \\}
 	\label{fig:cshkp} 
\end{figure}

The traditional CSHKP model of the two-ribbon eruptive flare 
\citep{Carmichael1964, Sturrock1966,Hirayama1974, Kopp1976}, shown in 
Figure~\ref{fig:cshkp}a, is able to explain many of the generic, large-scale 
observational properties of solar flares.
Several three-dimensional (3D) generalizations of the CSHKP scenario have 
been proposed in the form of cartoons \citep{Moore2001,Priest2002}, quantitative topological models \citep{Longcope2007}, and analytic flux rope solutions \citep{Isenberg2007}. 
Figure~\ref{fig:cshkp}b shows the \citet{Longcope2007} schematic of the 
CSHKP scenario where reconnection occurs at several sites to create the 3D coronal 
flare arcade loops and an erupting CME flux rope. 
Three-dimensional magnetohydrodynamics simulations of CME initiation all produce---to a greater or lesser extent---some version of the Figure~\ref{fig:cshkp}b eruptive scenario \citep[e.g.][]{Torok2005,Torok2011,Fan2007,Roussev2007,Lugaz2011,Aulanier2012,Lynch2009,Lynch2016b}. 

Figure~\ref{fig:cshkp} shows that one of the main properties 
characterizing solar flares is the amount of magnetic flux that reconnects. While reconnected flux cannot be measured directly from observations of the corona, the CSHKP model implies a quantitative relationship between the reconnection flux in the corona and the magnetic flux swept by the flare ribbon \citep[e.g.][]{Forbes1984} given by
\begin{equation} 
\frac{\partial \Phi}{\partial t}=\frac{\partial}{\partial t} \int B_c \, dS_c = \frac{\partial}{\partial t} \int B_\mathrm{n} \, dS_\mathrm{ribbon} . 
\label{drbneq} 
\end{equation} 
The left hand side, $\partial \Phi/\partial t$, denotes the coronal magnetic 
reconnection rate as {\it reconnection flux per unit time} defined by the 
integration of the inflow coronal magnetic field, $B_c$, over the reconnection area, $dS_c$. 
On the right-hand side, $B_\mathrm{n}$ is the normal component of the magnetic field 
in the ribbons which are the footpoints of the newly reconnected magnetic field 
lines in the corona. While direct measurements of $B_c$ and $dS_c$ in the corona 
are not currently feasible, $B_\mathrm{n}$ and $dS_\mathrm{ribbon}$ are relatively 
straightforward to obtain from photospheric magnetogram and lower-atmosphere 
flare ribbon observations. Summing the total normal flux swept by the flare ribbon 
area 
\begin{equation} 
\Phi_\mathrm{ribbon} = \int \left( \partial \Phi / \partial t \right) dt = \int B_\mathrm{n} dS_{\rm ribbon}
\label{rbneq} 
\end{equation} 
yields an 
indirect, but well-defined, measure of the amount of magnetic flux processed by 
reconnection in the corona during the flare.

A number of studies has investigated the relationship between 
various flare properties and properties of the resulting CME: e.g. UV and HXR 
emission with the acceleration of filament eruptions \citep{Jing2005, Qiu2010}, 
CME acceleration and flare energy release (\citealt{Zhang2001,Zhang2006}), GOES flare 
class, flare reconnection flux and the CME speed and flux content of the 
interplanetary CME \citep{Qiu2005, Qiu2007, Miklenic2009, Hu2014,Salas-Matamoros2015,Gopalswamy2017}. However, 
in most of these analyses, the underlying data for the flare ribbon properties were 
of limited accuracy and involved different sets of instruments that required 
time-consuming co-alignment, making systematic comparison of flare ribbon 
properties difficult for large numbers of events.

The launch of the \emph{Solar Dynamics Observatory} 
\citep[\emph{SDO};][]{Pesnell2012}, with the Helioseismic and Magnetic 
Imager \citep[HMI;][]{Scherrer2012,Hoeksema2014} and the Atmospheric 
Imaging Assembly \citep[AIA;][]{Lemen2012} instruments, represents the 
first time that both a vector magnetograph and ribbon-imaging capabilities 
are available on the same observing platform, making co-registration of 
AIA and HMI full-disk data relatively easy. 
In this paper, we present a database of flare ribbons associated with \nflares 
events corresponding to all flares of GOES class C1 and larger, with heliographic longitudes less than 45 degrees, from April 2010 through April 2016. Our intentions are twofold. First, we provide the reference for the dataset by describing the key processing procedures.  Second, we present the statistical analyses of the flare reconnection fluxes and their relationship with other flare and AR properties.

This is the first in a series of two papers. Here, we focus on the cumulative reconnection properties, while in the second paper, we will analyze their temporal evolution.


This paper is organized as follows. In section~\ref{methods}, we describe 
the \emph{SDO} data and the analysis procedure for correcting pixel 
saturation, creating the flare ribbon masks, and calculating the ribbon 
reconnection fluxes and their uncertainties. In section~\ref{data}, we summarize 
the database of events, describe the AR and flare ribbon properties 
calculated for each of our events, compare these to the flare GOES peak 
X-ray fluxes, and present the distribution of the magnetic energy estimates 
associated with the reconnection fluxes. In section~\ref{disc}, we discuss our 
results, and in section~\ref{conc}, summarize our conclusions.

\section{Data \& Methodology}\label{methods}

In this section, using an X2.2 flare in NOAA AR 11158 as an example, we describe how we correct the AIA 1600\AA{} saturated pixels (section \ref{subsec:pixsat}), identify the flare ribbons and find the reconnection fluxes (section~\ref{subsec:tempevol}). 
 
\subsection{Filtering the Pixel Saturation in AIA 1600\AA{} Observations}
\label{subsec:pixsat}
\begin{figure*}[htb!]
 \begin{center}
 \includegraphics[width=0.9\textwidth]{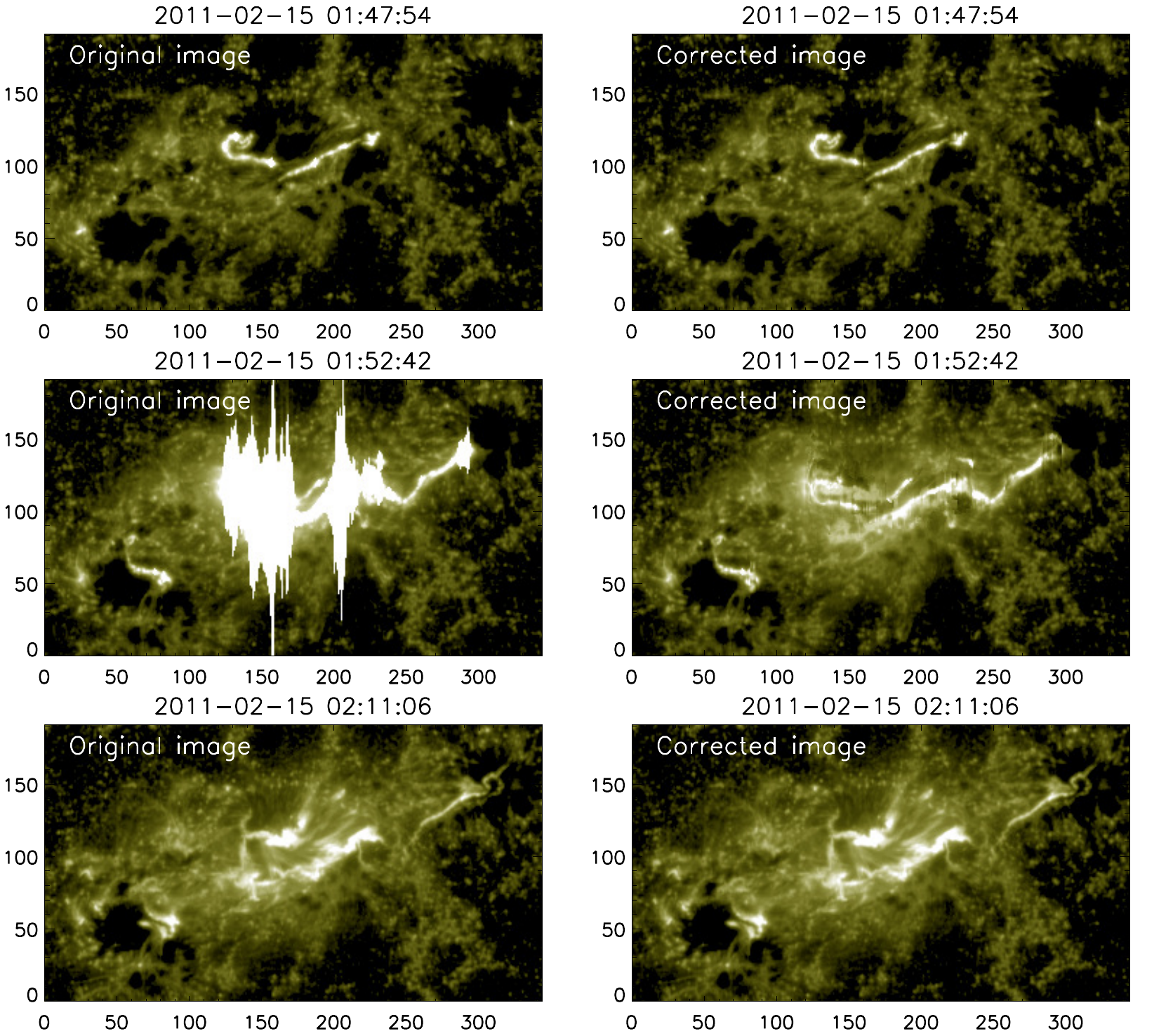}
 \end{center}
 \caption{{Snapshots of the 1600\AA{} flare ribbons in the 
 	X2.2 flare in NOAA AR 11158 observed by AIA on 2011 February 
	15.
	Left column: original AIA image sequence. Right column: 
	saturation-corrected image sequence. Top row: no saturated pixels 
	at the very beginning of the impulsive phase. Middle row: maximum 
	number of saturated pixels. Bottom row: no saturated pixels during 
	the gradual phase of the flare. \\}}
 \label{Figure:aia0} 
\end{figure*}

The key technical challenge of defining the set of pixels corresponding 
to the flare ribbon location, $dS_\mathrm{ribbon}$, in the AIA image sequences 
is the correction of saturated pixels caused by CCD saturation, pixel 
bleeding, and of the diffraction patterns from the EUV-telescope entrance 
filter. Unfortunately, existing software packages for automatic de-saturation 
of AIA images such as DESAT \citep{Schwartz2015} are not applicable to 
the 1600\AA{} channel (Gabriele Torre, private communication). Here we 
present our own empirical approach to correct the intensities of ``bloomed'' 
pixels.

To describe the details of our saturation-correction approach we use the 
\emph{SDO} AIA observations of the well-known ``Valentine's Day" flare as 
a representative example. 
%
%
This flare occurred in NOAA AR 11158 on 2011 February 15, 01:44 UT 
\citep{Schrijver2011}. The \emph{SDO} AIA observations of this event were 
saturated during the impulsive phase, from 01:49 UT to 02:10 UT in 
the UV 1600\AA{} continuum as well as in other AIA bands. We re-examine 
this event in UV 1600\AA{} observations with 24-second cadence and 
$0.\arcsec 61$ pixel resolution with the objective of 
removing the saturated pixels and reconstructing the evolution of the UV 
ribbons from the earlier and later (unsaturated) phases of the flare. We 
process the UV 1600\AA{} images in IDL using the \verb+aia_prep.pro+ 
SolarSoft package and co-align the AIA image sequence in time with the 
first frame.
 
Our saturation-correction approach includes the following steps. We first 
select the pixels above saturation level, $I_{sat}=5000$ counts~s$^{-1}$, and 
pixels surrounding them within 2 and 10 pixels in the $x$- and 
$y$-directions. We then replace each saturated pixel intensity with the value linearly 
interpolated \emph{in time} between the individual pixel's previous and subsequent 
unsaturated values that bracket the saturation duration.
Figure~\ref{Figure:aia0}, left column, shows a sequence of original AIA 
1600\AA{} images on 2011-02-15: top panel, before the impulsive phase 
when AIA observations had no saturated pixels (01:47UT); middle panel, 
at the peak of the impulsive phase with the largest number of saturated 
pixels (01:52UT); and lower panel, during the gradual phase with no 
saturated pixels (02:11UT). Figure~\ref{Figure:aia0}, right column, shows 
the saturation-corrected images which differ from the original images only 
in the location of the saturation-corrected pixels. This empirical approach, while not suitable for photometric analysis of the corrected images, does allow one to identify flare ribbon locations (compare original and corrected panels of the middle row). 
Thus the saturation-corrected 1600\AA\ image sequence provides sufficient 
information to determine reconnected flux using Equation~\ref{rbneq}.

\begin{figure}
 \centerline{ \includegraphics[width=4.0in,angle=0]{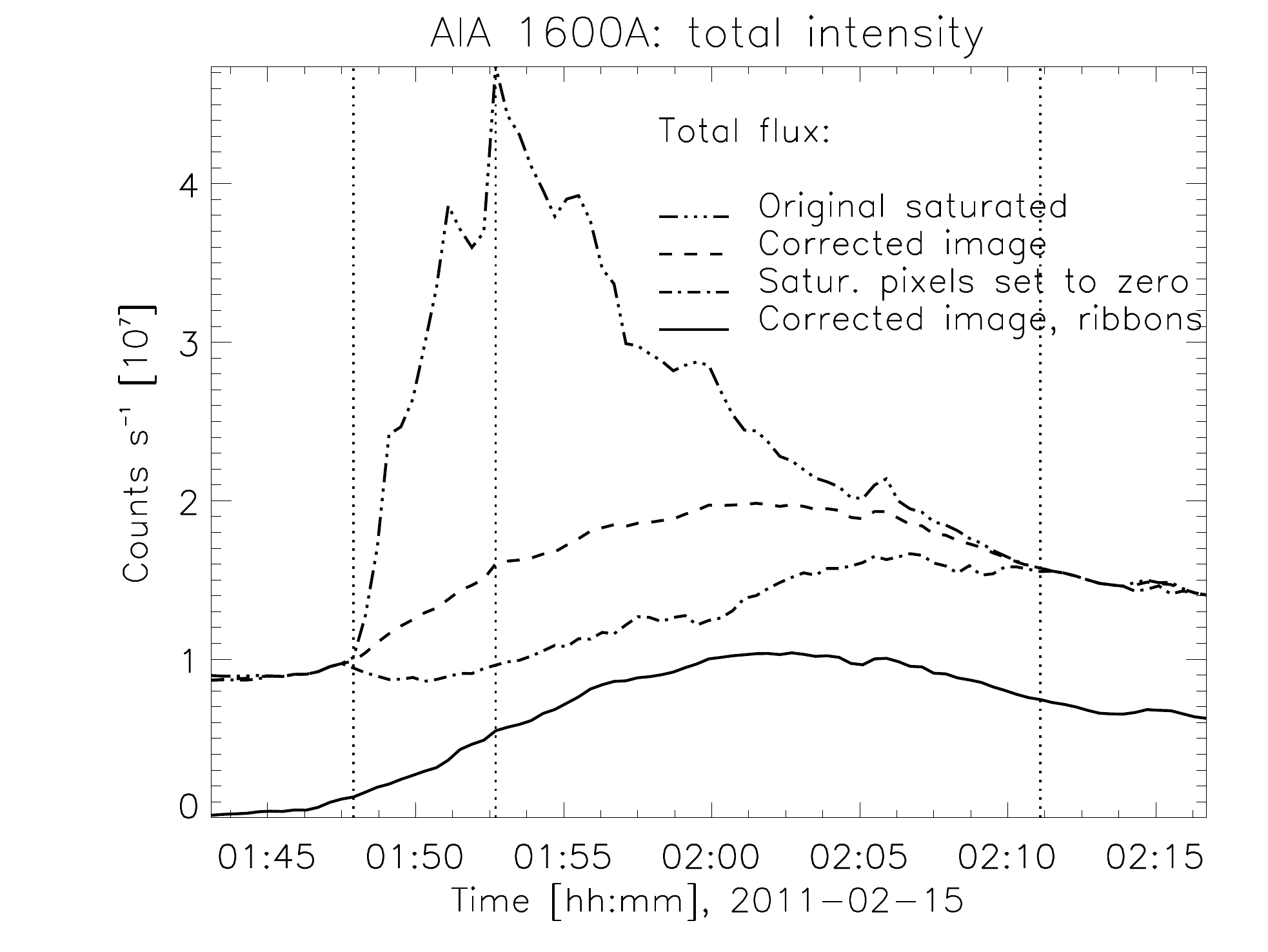} }
 \caption{Area-integrated light curves of the AIA 1600\AA{} at different steps of the saturation removal procedure. Solid line shows 
	counts in the ribbons alone, i.e. pixels above $c=8$ times the background median value. Three vertical dotted lines correspond to three rows in Figure~\ref{Figure:aia0} \\ }
 \label{Figure:aiaflx} 
\end{figure}

Figure~\ref{Figure:aiaflx} shows the area-integrated light curves of the 
AIA 1600\AA{} image sequence at each step of the saturation removal 
procedure. The dash-triple-dot curve, labeled `Original saturated,' 
plots the total number of counts in the saturated AIA 1600\AA{} image 
sequence. The period from 01:49 to 02:10UT, shown with vertical dotted 
lines, indicates the duration of the pixel saturation in the sequence. 
The dashed curve, labeled `Saturated pixels set to zero,' plots the total 
number of counts after the saturated pixels above the threshold level of 
$I_{sat}$ and the adjacent pixels have been removed. 
The dash-dot curve, labeled `Corrected image', plots the total counts after 
the saturated pixel values have been replaced by the interpolation in time 
between the unsaturated values of those pixels in the image sequence. 
Finally, the solid curve, labeled `Corrected image, ribbons,' plots the 
light curve of the flare ribbons alone --- the set of pixels that are identified by 
the ribbon search algorithm described in the next section. Note that the `Corrected 
image, Ribbons only' light curve is offset from the `Corrected image' curve 
by the total flux of the background that remains nearly constant
during the flare. To summarize, Figure~\ref{Figure:aiaflx} quantitatively describes
how much of the original image intensity is affected by the saturation and what fraction
of this intensity is attributed to the flare ribbons using our saturation-correction approach.

\begin{figure*}[!htb]
 \begin{center}
 \includegraphics[width=0.99\textwidth]{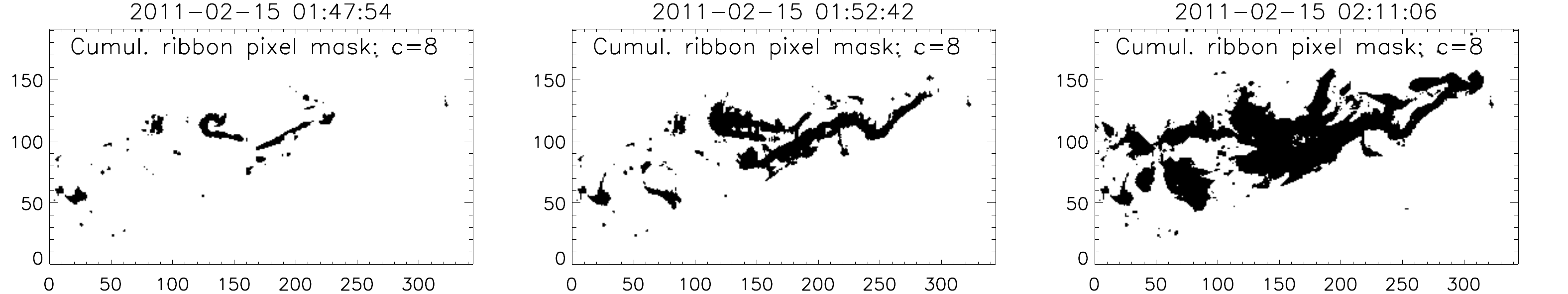}
 \end{center}
 \caption{{The 1600\AA{} cumulative ribbon pixel mask $M^\mathrm{(I_8)}(x_i,y_j,t_k)$ 
 	evolution during the 2011 February 15 X2.2 flare. 
	}}
 \label{Figure:mxyt} 
\end{figure*}

\subsection{Constructing Flare Ribbon Masks and Calculating the Reconnection Fluxes}
\label{subsec:tempevol} 

To find the reconnection flux as defined by Equation~\ref{rbneq} we need to know the flare ribbon location and the normal component of the magnetic field.

To identify the flare ribbon locations we use the AIA 1600\AA\ saturation-corrected image sequence from Section~\ref{subsec:pixsat} and the methodology of \citet{Qiu2002,Qiu2004,Qiu2007}.  
We define an {\it instantaneous} flare ribbon pixel mask $N^\mathrm{(I_c)}(x_i,y_j,t_k)$ in each pixel $(x_i,y_j)$ at each time step $t_k$  in the sequence with a value of one if the 1600\AA\ intensity is greater than an empirical  {\it ribbon-edge cutoff intensity}, $I_\mathrm{c}$, and with a value of zero if the  intensity is below  $I_\mathrm{c}$.
%
The cutoff intensity $I_\mathrm{c}$ for identifying the flare ribbon pixels ranges from the cutoff threshold $c=6$ to $c=10$ times the median image intensity at each time $t_k$.  This range is consistent with the range previously used for TRACE 1600\AA{} UV data \citep{Kazachenko2012}. Since the cutoff threshold for the ``steady-state" UV brightening associated with plage regions is typically $c \approx 3.5$  \citep[see Figure 3 of][]{Qiu2010}, our empirical  threshold range, $c\in[6,10]$, is significantly greater than typical non-flare related UV emission and is appropriate to capture the flare ribbons.

To find the normal component of the magnetic field, $B_n$, we use the full disk HMI vector magnetogram data series (\verb+hmi.B_720s+) in the form of field strength, inclination, and azimuth in the plane-of-sky coordinate \citep{Hoeksema2014}. We perform a coordinate transformation and decompose the magnetic field vectors into three components in spherical coordinates\footnote{Derivation of the radial component of the magnetic field is performed using the HMI pipeline code that is available to the public through the SDO webpage. Examples of usage can be found at \url{http://jsoc.stanford.edu/data/hmi/ccmc/}.} \citep{Sun2013}. The derived radial component is the normal component $B_n$ that we need in Equation~\ref{rbneq}. To avoid noisy magnetic fields, we only use magnetic fields greater than the flux density threshold $| B_{n} | > 100$~G \citep[see Fig.~2 in][]{Kazachenko2015}. 

The {\it unsigned reconnection flux} or unsigned magnetic flux swept up by flare ribbons up to time $t_k$ is then calculated using the discrete observations as
\begin{eqnarray}
\Phi_\mathrm{ribbon}^\mathrm{(I_c)} (t_k) = \int \limits_\mathrm{I_c} |B_\mathrm{n}(t_k)| \ dS(t_k)\approx \\
\approx \sum \limits_{i,j} |B_\mathrm{n}(x_i,y_j,t_{HMI})| \ M^\mathrm{(I_c)}(x_i,y_j,t_k) \ ds^2_{ij} \,
\end{eqnarray}
where $t_\mathrm{HMI}$ corresponds to the time of the measurement of the normal component of the magnetic field $B_n$ and the ribbon area $dS(t_k)$ above the ribbon-edge cutoff intensity, $\mathrm{I_c}$, is given by the discrete {\it cumulative}
ribbon pixel mask $M^\mathrm{(I_c)}(x_i,y_j,t_k)$ multiplied by the pixel area $ds_{ij}^2$. We correct each individual pixel area for foreshortening: $ds^2_{ij}={ds^2_{ij,{\rm obs}}}/{\cos{\theta_j}}$, where $\theta_j$ is the angular distance between the central meridian and the pixel $(x_i, y_j)$. The cumulative ribbon pixel mask $M^\mathrm{(I_c)}(x_i,y_j,t_k)$ is related to the instantaneous ribbon pixel mask 
$N^\mathrm{(I_c)}(x_i,y_j,t_k)$ in the following way
\begin{equation*}
M^\mathrm{(I_c)}(x_i,y_j,t_k) = M^\mathrm{(I_c)}(x_i,y_j,t_{k-1}) \cup N^\mathrm{(I_c)}(x_i,y_j,t_k).
\end{equation*}
Thus, the cumulative ribbon pixel mask represents the time integral (accumulation) of every flare ribbon pixel 
$(x_i, y_j)$ that exceeds the 1600\AA\ intensity threshold at some instance from the first frame of the image sequence to $t_k$.
Figure~\ref{Figure:mxyt} shows the cumulative ribbon pixel mask $M^\mathrm{(I_8)}(x_i,y_j,t_k)$ consisting of pixels above $c=8$ times the background median value from the sequence of corrected images at the times shown in Figure~\ref{Figure:aia0}. 

\begin{figure*}[htb!]
	\begin{center}
	 \centerline{ 
	\includegraphics[width=0.64\textwidth]{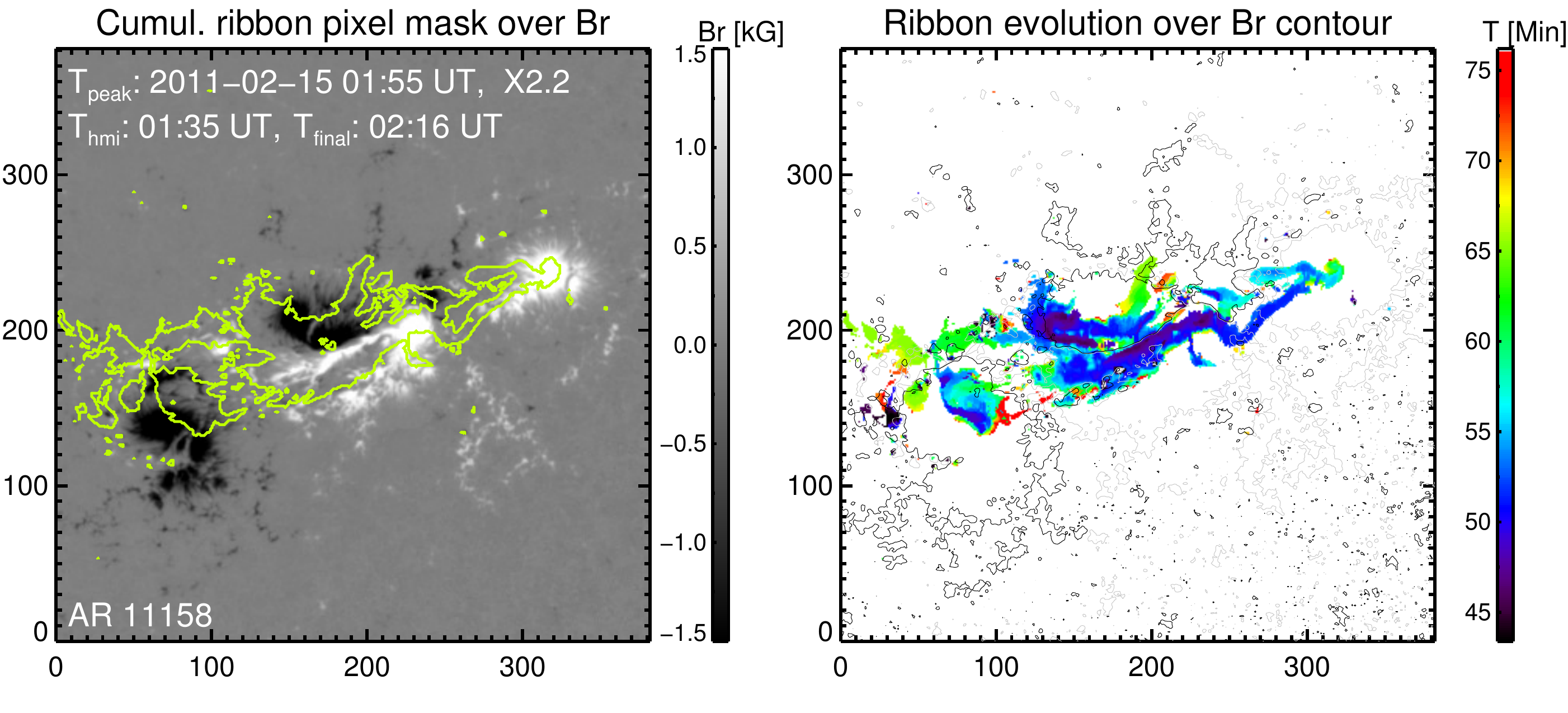} 
	\includegraphics[width=0.36\textwidth]{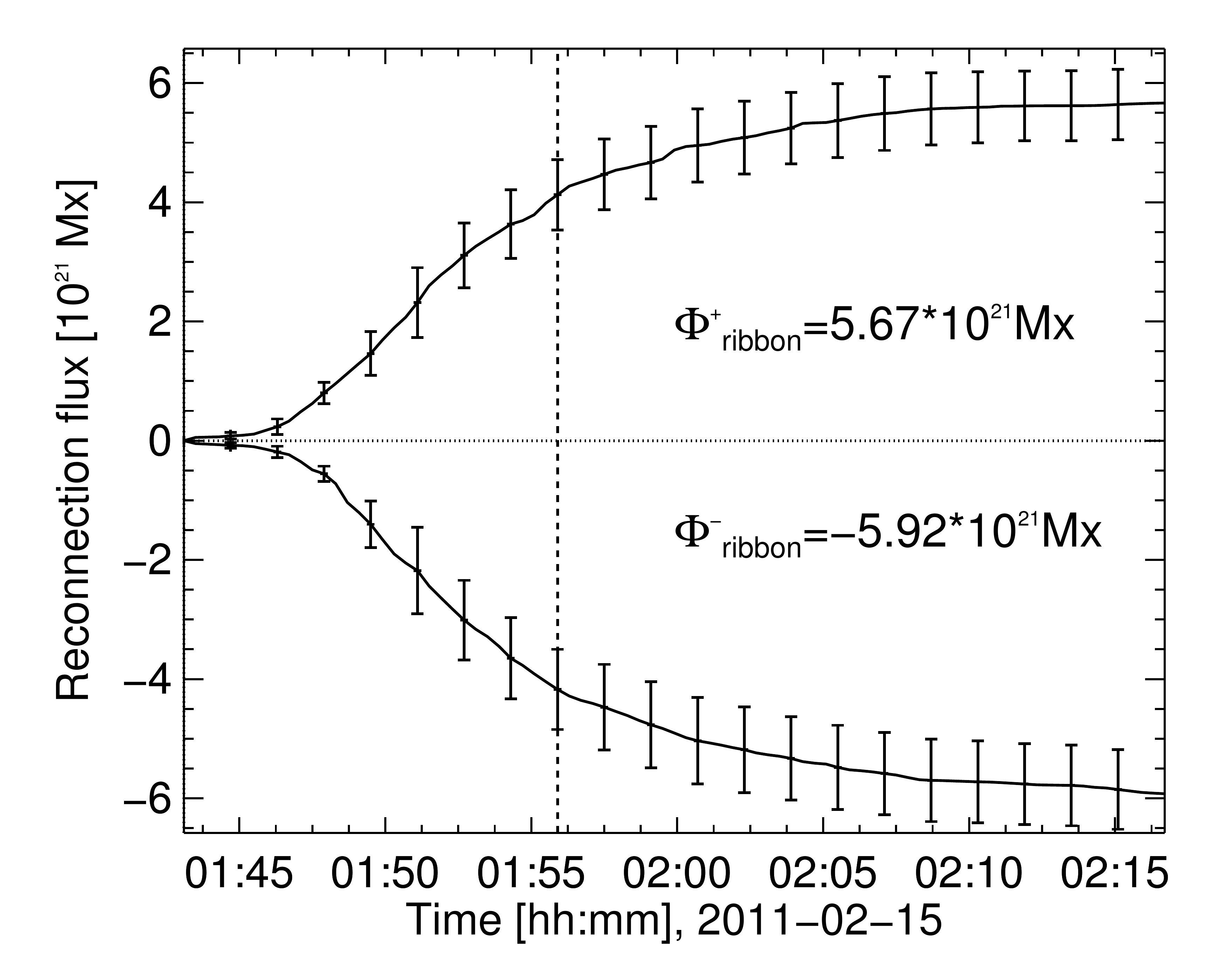} 
	}
 	\caption{{\emph{Left}: The HMI photospheric magnetogram $B_\mathrm{n}$ 
	with the contours of the cumulative AIA 1600\AA{} flare ribbons at the flare end time over-plotted. The times in the top left corner are: GOES peak X-ray flux time ($t_\mathrm{peak}$), time of HMI $B_\mathrm{n}$ observation ($t_\mathrm{hmi}$), and the flare end time ($t_\mathrm{final}$). \emph{Middle}: temporal and spatial evolution of the UV flare ribbons 
	$M^\mathrm{(I_8)}(x,y,t_{\rm final})$ with each pixel colored by the time of its initial brightening. 
	\emph{Right}: Time profiles of the total reconnection flux in units of Mx integrated in the positive and negative polarities, 
	respectively. The error bars in the reconnection flux indicate the range of uncertainty from the ribbon area 
	identification. Vertical dotted line marks GOES peak X-ray flux time. $\Phi^+_\mathrm{ribbon}$ and $\Phi^-_\mathrm{ribbon}$ indicate positive and negative reconnection fluxes at the end of the sequence at time $t_\mathrm{final}$.}}
 \label{Figure:aiapos} 
\end{center}
\end{figure*}

In Figure~\ref{Figure:aiapos} we summarize the temporal evolution of flare ribbons associated with an X2.2 flare on 2011 Feb 15. The left panel shows the contours of the maximum flare ribbon area at the end of the AIA image sequence ($t_{\rm final}$) 
superimposed on the co-aligned HMI magnetogram before the flare at $t_{\rm hmi}$. The middle panel shows the temporal evolution of the ribbons, with blue and red colors corresponding to early and late stages of the flare respectively. Note that the 
$M^\mathrm{(I_8)}(x_i,y_j,t_k)$ color-coded bitmaps are plotted in reverse-temporal order so that every individual pixel in the cumulative ribbon pixel mask is colored according to the time of its \emph{initial} brightening. 

The right panel of Figure~\ref{Figure:aiapos} shows the evolution of magnetic fluxes swept up by ribbons in positive and negative polarities, the {\it signed reconnection fluxes}, $\Phi^+_\mathrm{ribbon}$ and $\Phi^-_\mathrm{ribbon}$, respectively.
To reflect the uncertainty in these estimates due to the ribbon area identification, we perform the entire pixel mask area calculation twice: once with the cutoff threshold $c=6$, and again with $c=10$. 
Then, the signed reconnection fluxes in each polarity at time $t_k$ are
\begin{align}
\Phi^{+}_\mathrm{ribbon}= \frac{{\Phi^{+(I_6)}_{\rm ribbon}} + \Phi^{+(I_{10})}_{\rm ribbon}}{2}, \\
\Phi^{-}_\mathrm{ribbon} =\frac{{\Phi^{-(I_6)}_{\rm ribbon}} + \Phi^{-(I_{10})}_{\rm ribbon}}{2},
\end{align}
where ``$+$'' and ``$-$'' refer to integration over positive and negative polarities, respectively. 
In Figure~\ref{Figure:aiapos} the reconnection fluxes in both polarities evolve nearly simultaneously. By $t_{\rm final}=$~02:16UT, the positive and negative reconnection flux is 
$\Phi^+_{\mathrm{ribbon}} \ (\Phi^-_{\mathrm{ribbon}}) = 5.67 \ (-5.92) \times 10^{21}$~Mx, and the total unsigned reconnection flux is $\Phi_{\mathrm{ribbon}}=|\Phi^{+}_\mathrm{ribbon}|+|\Phi^{-}_\mathrm{ribbon}|=1.16\times10^{22}$~Mx. 
Theoretically, equal amounts of positive and negative flux should be reconnected. Hence, the balance between the two increases the credibility of the applied technique. 

We estimate the errors in $\Phi^{+}_\mathrm{ribbon}$ and $\Phi^{-}_\mathrm{ribbon}$ at time $t_k$ using the uncertainty in the ribbon area (see error bars in Figure~\ref{Figure:aiapos}):
\begin{align*}
\Delta \Phi^{+}_\mathrm{ribbon} = \frac{{\Phi^{+(I_6)}_{\rm ribbon}} - \Phi^{+(I_{10})}_{\rm ribbon}}{2}, \\
\Delta \Phi^{-}_\mathrm{ribbon} = \frac{{\Phi^{-(I_6)}_{\rm ribbon}} - \Phi^{-(I_{10})}_{\rm ribbon}}{2}.
\end{align*}
Typically these range within $10\%$ to $20\%$ of $\Phi^{\pm}_\mathrm{ribbon}$. Further in text we do not take into account the uncertainty associated with the physical height of ribbon formation. The 1600\AA{} UV emission corresponds to the upper chromosphere and transition region whereas the photospheric magnetic field measurements 
are $\sim$2~Mm below this. The differences in normal field strength at the photosphere and at the ribbon formation height typically lead to a maximum decrease in the total reconnection flux of 10--20\% \citep{Qiu2007, Kazachenko2012}.

\section{Database Description}\label{data}	
\subsection{Flare Ribbon Event List}\label{ssFREL}
\begin{figure*}[!tb]
	\includegraphics[width=1.0\textwidth]{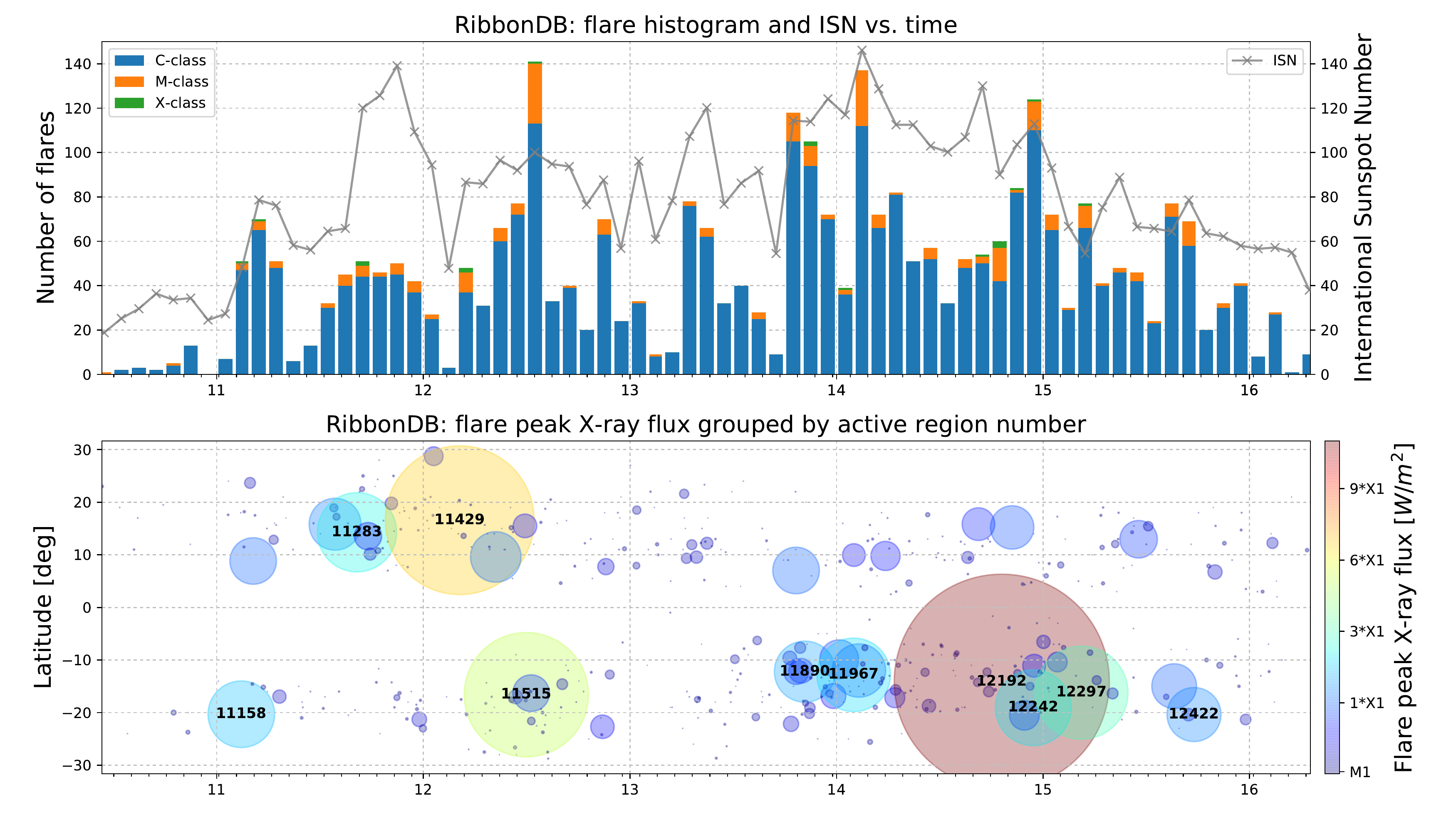} 
	\caption{{{\it Top panel:} Number of C-, M- and X-class flares each month 
	in the flare-ribbon database {\tt RibbonDB} (left axis) and sunspot number from April 2010 
	until April 2016 (right axis); {\it Bottom panel:} flare peak X-ray flux and location 
	on the disk grouped by AR versus time: circle size and color correspond 
	to the peak X-ray flux summed over each AR number. \\}}
 \label{fig:ssn} 
 \end{figure*}

The flare ribbon catalogue {\tt RibbonDB} contains properties of the ARs 
and flare ribbons for all well-observed flares of GOES class C1.0 and larger in the 
\emph{SDO} era, from April 2010 until April 2016. We used the existing Heliophysics Event Catalogue (HEC) 
maintained by the INAF-Trieste Astronomical Observatory to select the events for our flare 
ribbon catalogue. We chose flares within $45^\circ$ from the central meridian to minimize projection effects. We also excluded 
events that were missing the AR number. We used \verb+get_nar.pro+ from \verb+SolarSoft+ to obtain the AR location coordinates for events with missing 
AR coordinates. These criteria resulted in \nflares flares for our {\tt RibbonDB} 
catalogue, including 17 X-class, 250 M-class, and 2870 C-class flares (see Table~\ref{flares}). Each entry contains the following information from the HEC: flare start time, flare peak time, flare 
end time, flare peak X-ray flux (flare class), flare heliographic longitude and latitude, 
and AR number.
\begin{table}[htb!]
	\caption{Number of C, M, and X flares and their corresponding percentages in the {\tt RibbonDB} catalogue.} 
	\begin{center}
	\begin{tabular}{rrr}
\toprule
\bf{ Class} 		& \bf{ Number of flares} 			& \bf{ Percentage, $\%$} \\
 \hline
C		& $2870$		& $91.5$ \\ 
M		& $250$		& $8.0$ \\
X		& $17$		& $0.5$ \\
\midrule
T		& $\nflares$		& $100.0$ \\
\bottomrule
\end{tabular}
\label{flares} 
\end{center} 
\end{table}
Figure~\ref{fig:ssn} shows monthly international sunspot number and number of C-, M-, and X-class 
flares selected for the \verb+RibbonDB+ catalogue as a function of time (upper panel) and 
also the time and location of ARs (lower panel). {\tt RibbonDB} covers the first half of solar cycle $24$, including its maximum around April $2014$. 
In the lower panel of Figure~\ref{fig:ssn}, the radius of each AR circle is 
proportional to the cumulative peak X-ray flux over the AR's lifetime.
The ARs with the largest cumulative peak X-ray fluxes are NOAA AR 12192 (24 
flares equivalent by the peak X-ray flux to 9~X1 flares), NOAA AR 11429 (13 flares equivalent to 6~X1 
flares), and NOAA AR 11515 (26 flares equivalent to 3~X1 flares).

\begin{figure*}
 	\centerline{ 
	\begin{tabular}{ll}
	\includegraphics[width=0.64\textwidth]{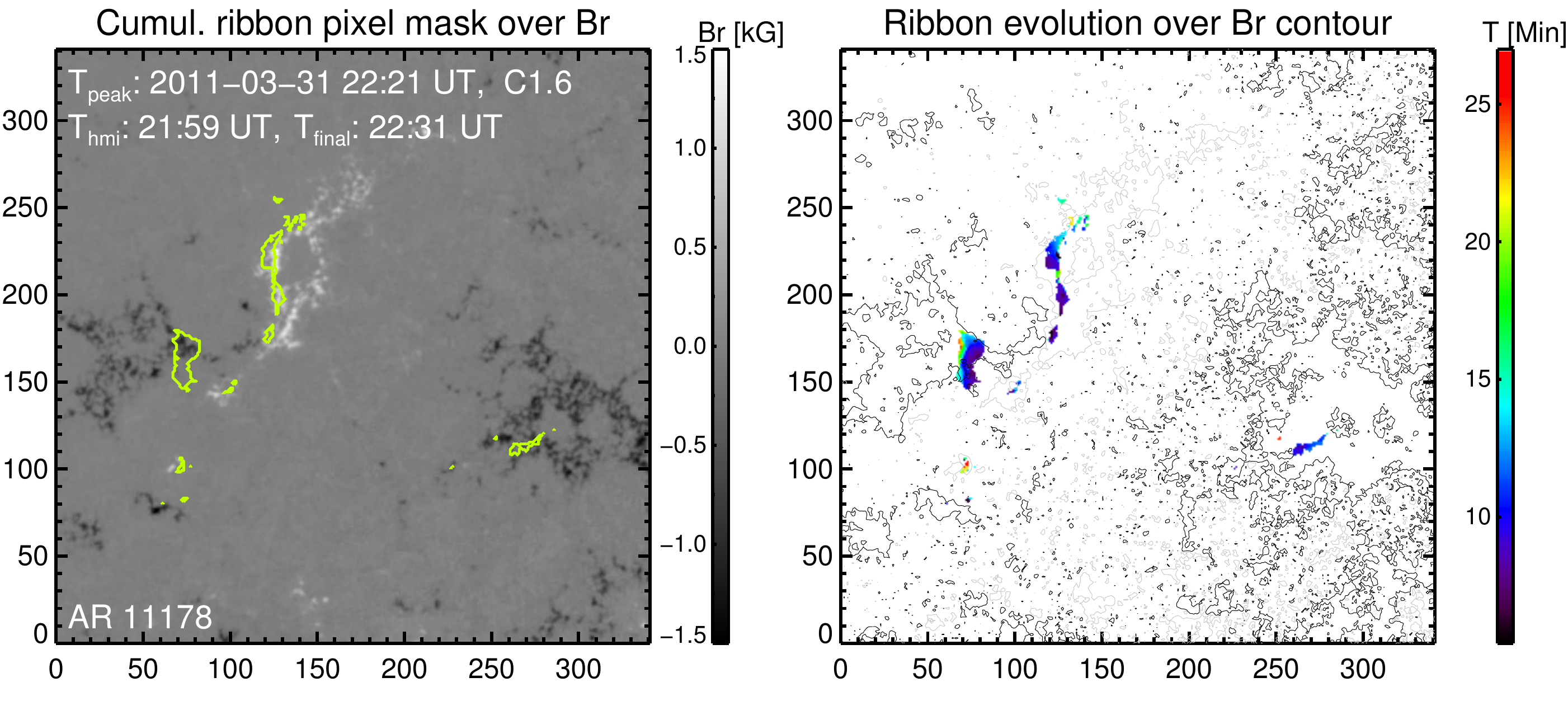}& 
	\includegraphics[width=0.36\textwidth]{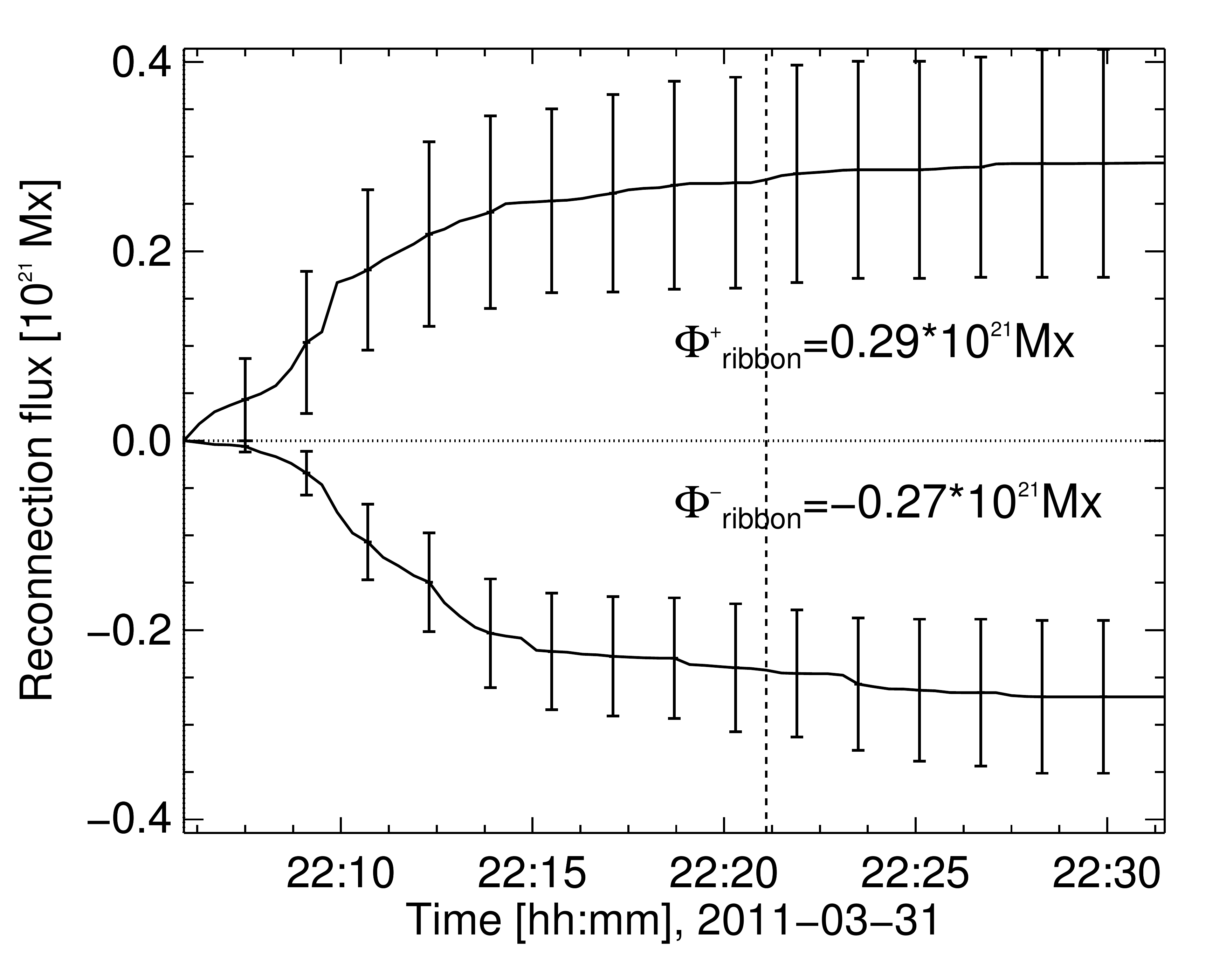}\\ 
	\includegraphics[width=0.64\textwidth]{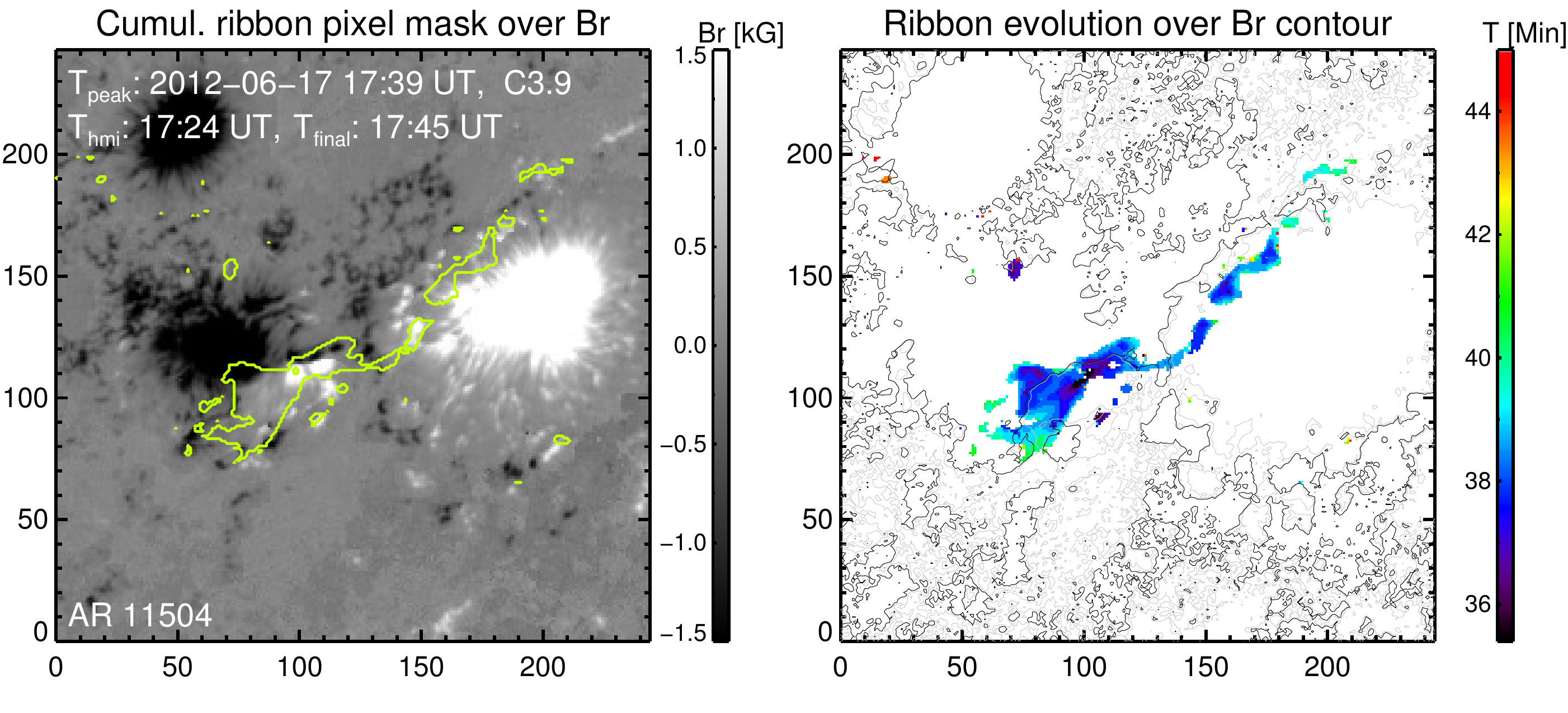}& 
	\includegraphics[width=0.36\textwidth]{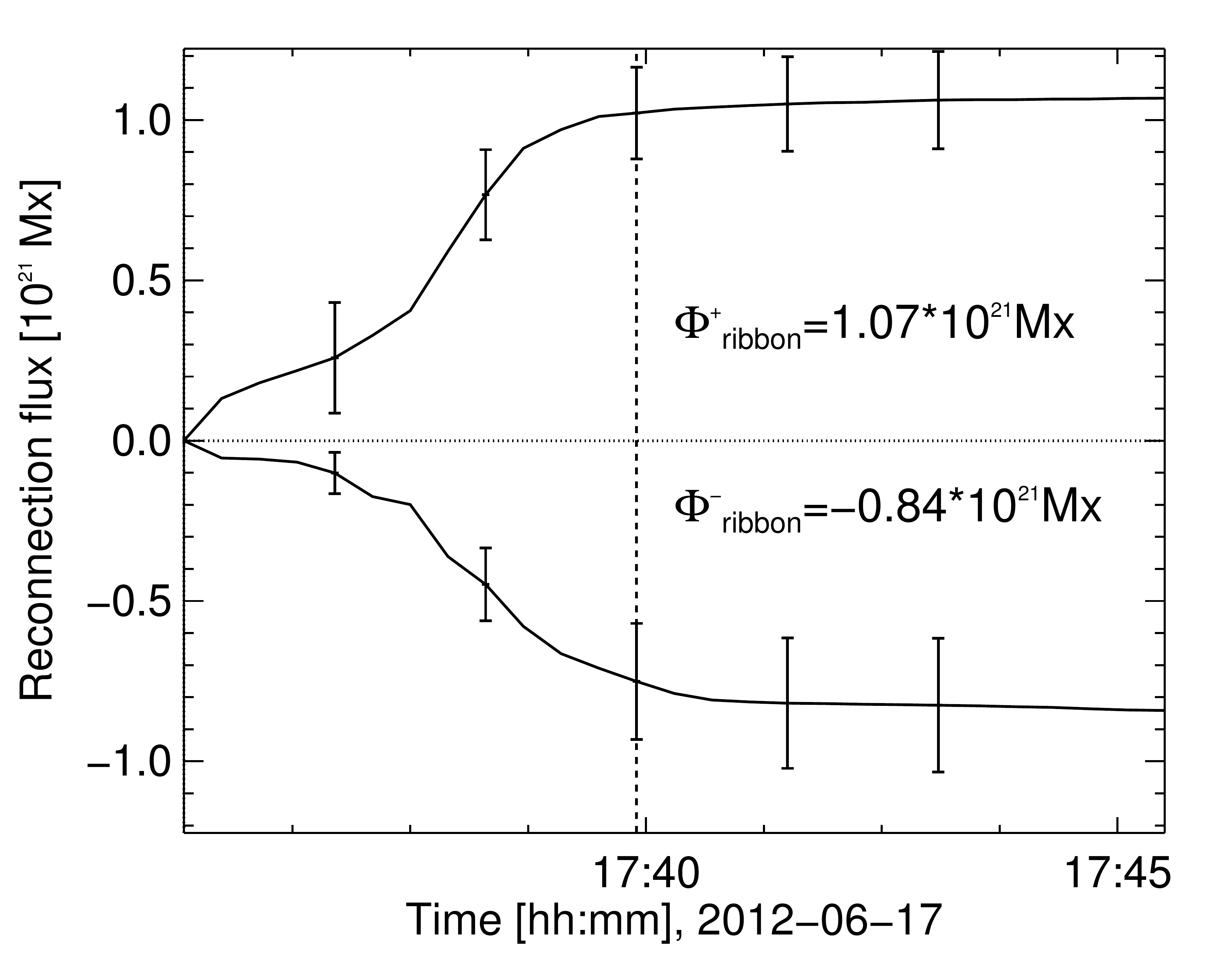}\\ 
	\includegraphics[width=0.64\textwidth]{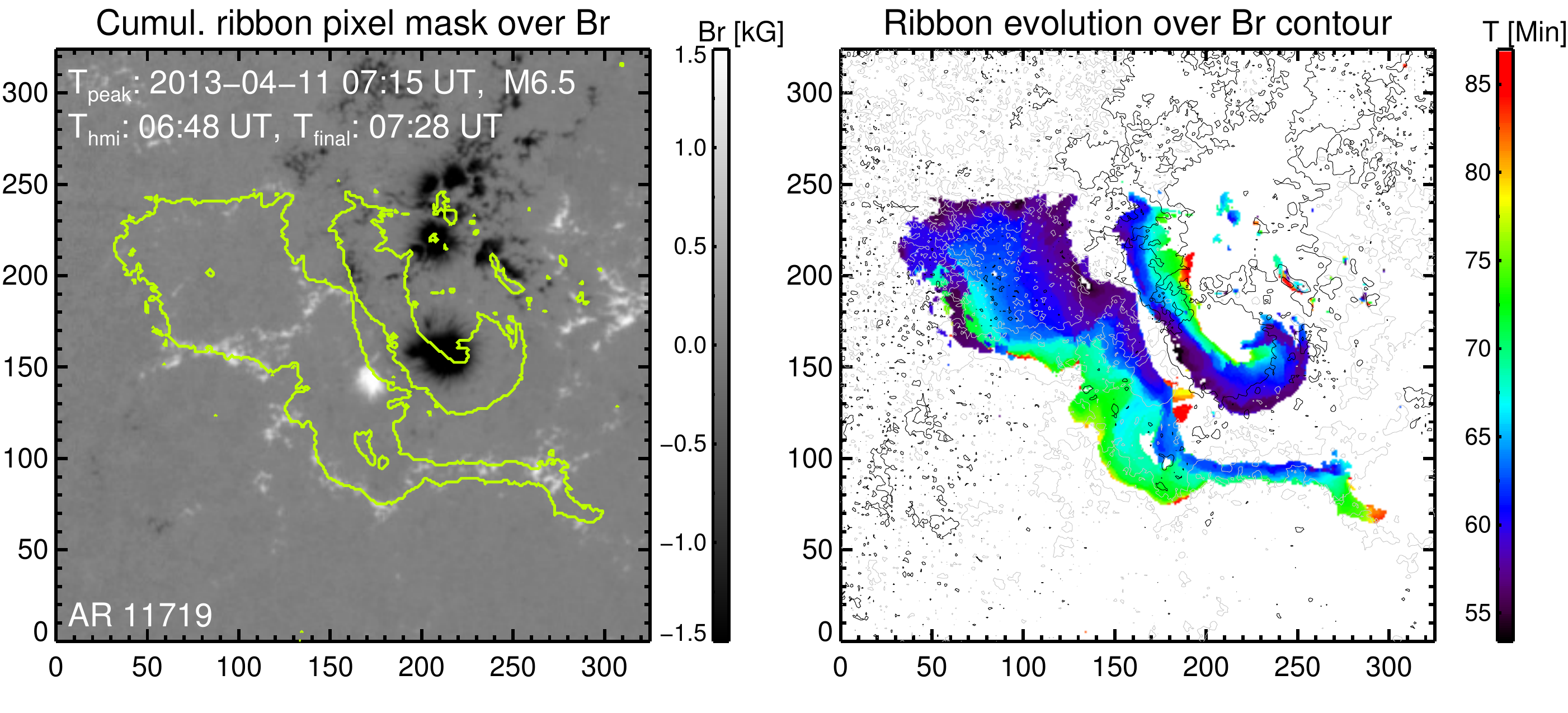}& 
	\includegraphics[width=0.36\textwidth]{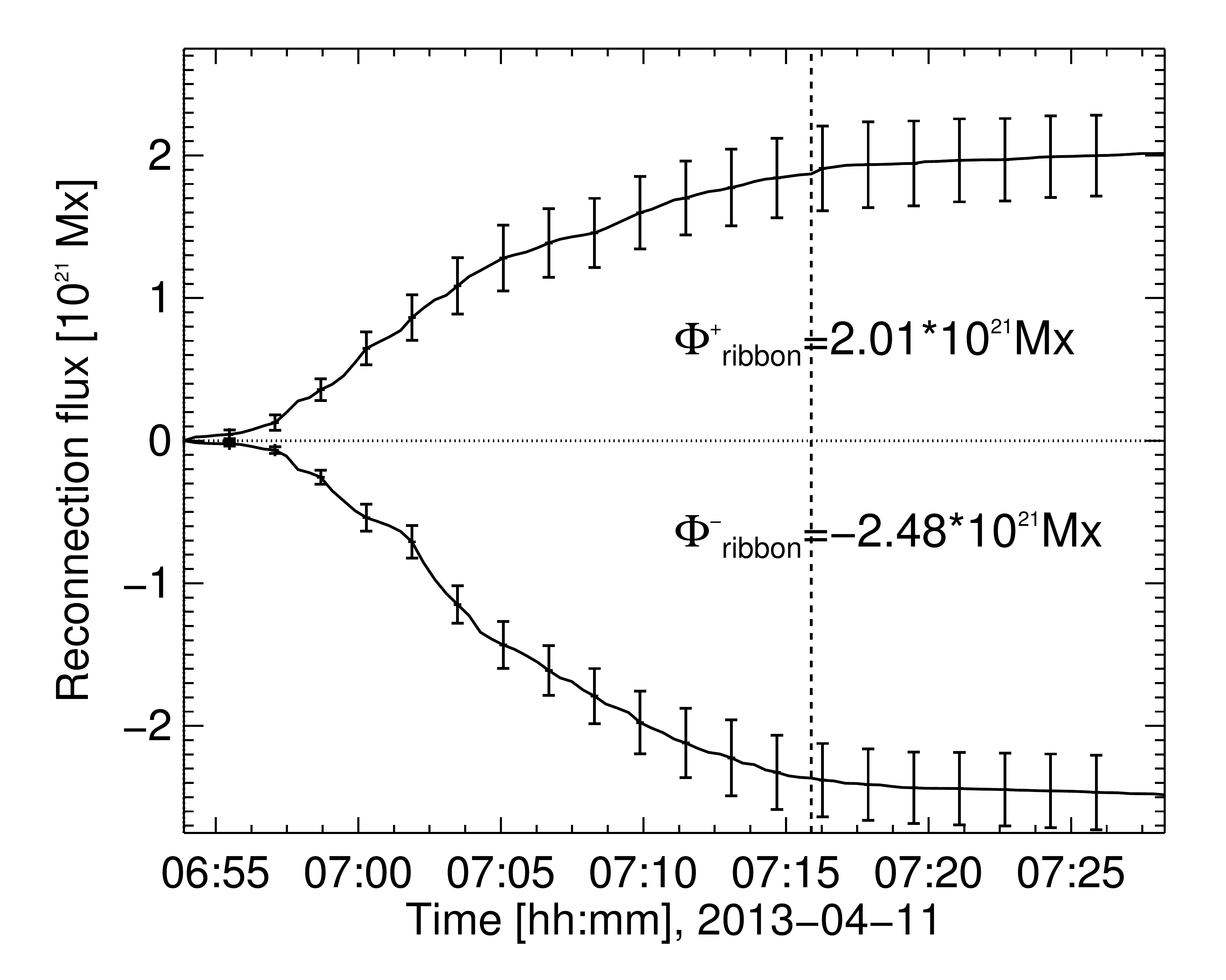}\\ 
	\includegraphics[width=0.64\textwidth]{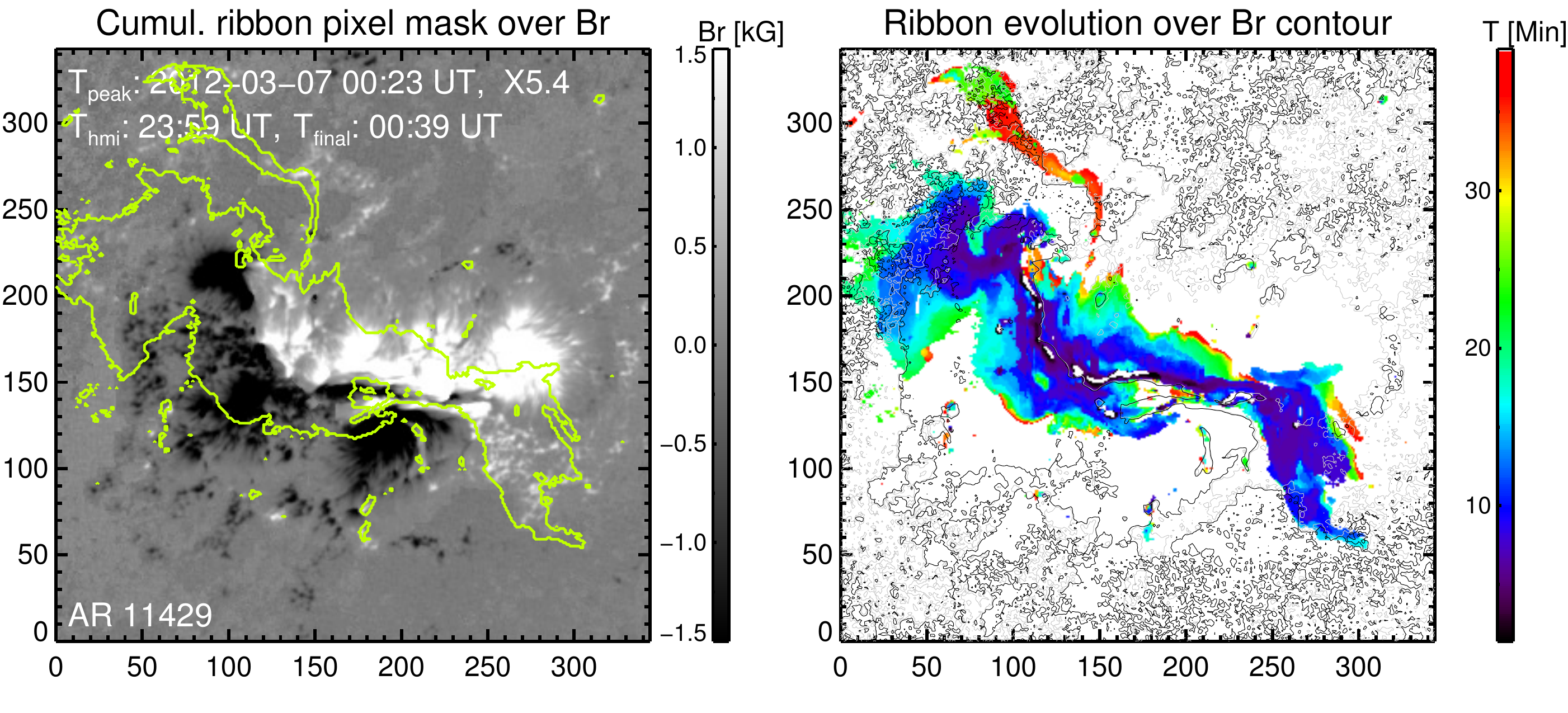}& 
	\includegraphics[width=0.36\textwidth]{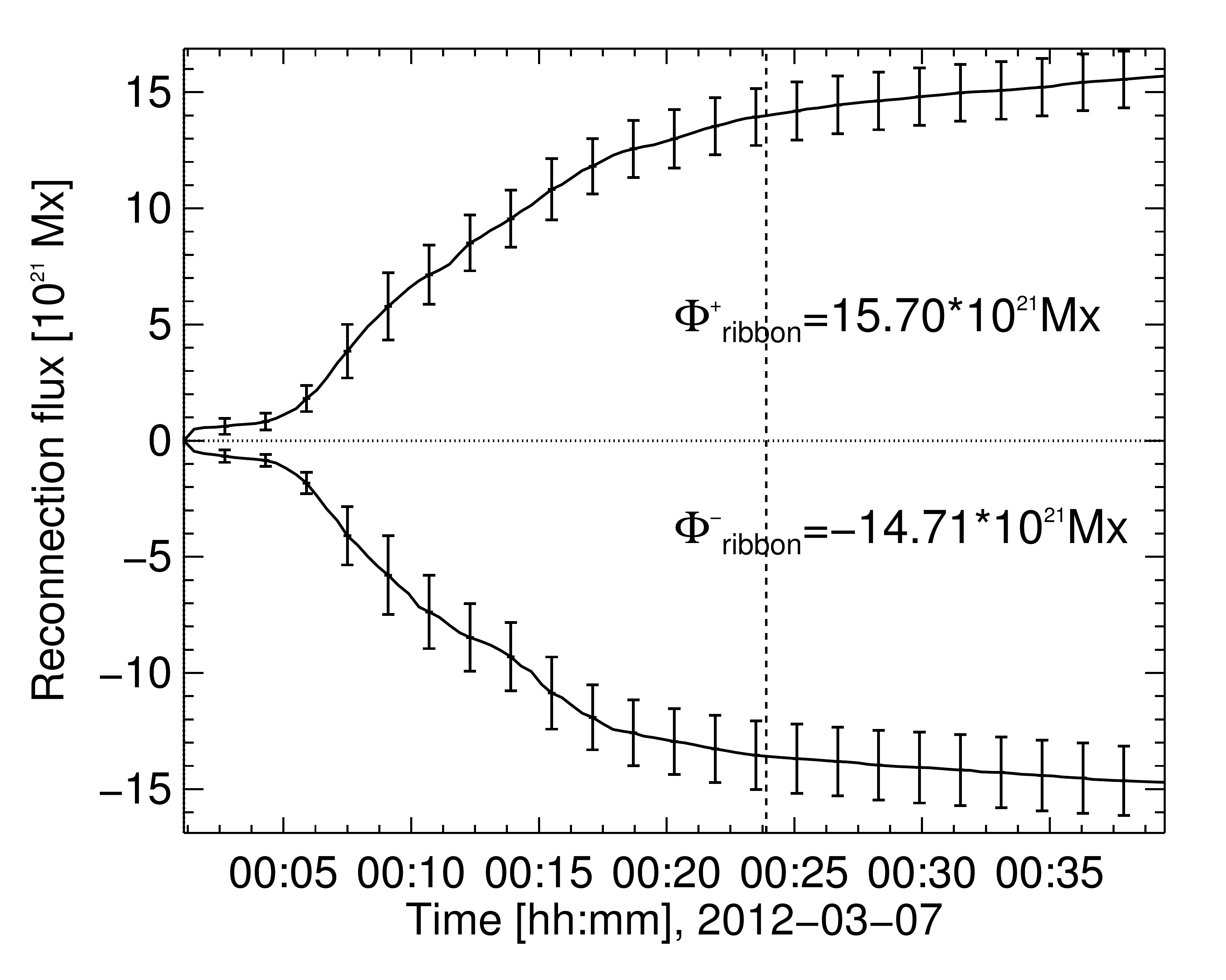}\\ 
	 \end{tabular}
	}
 	\caption{{Representative flare ribbon events from the database 
	in the same format as Figure~\ref{Figure:aiapos}. From top-to-bottom, the flare classes and GOES X-ray flux peak times are C1.6 on 
	2011-03-31 at 22:21UT, C3.9 on 2012-06-17 at 17:39UT, M6.7 on 
	2013-04-11 at 07:15UT, and X5.6 on 2012-03-07 at 00:23UT. }}
 \label{fig_fullpage} 
 \end{figure*}

Figure~\ref{fig_fullpage} shows four representative events from our {\tt RibbonDB} catalogue, ranging from GOES C1.6 to X5.6-class, in the same format as Figure~\ref{Figure:aiapos}.  

\subsection{AR and Flare Ribbon Properties}\label{dbanal}
For each event in the database we use the AIA $1600$\AA{} image sequence, of 24-seconds cadence and $0.61\arcsec$ spatial resolution, and a pre-flare HMI vector field magnetogram, of 12-minutes cadence and $0.5\arcsec$ spatial resolution, to derive the flare-ribbon-pixel masks and the normal component of the magnetic field, $B_\mathrm{n}$, accordingly (see Section~\ref{subsec:tempevol} for details).\footnote{As the beginning and the end times of the AIA $1600$\AA{} image sequence we chose one minute before the flare start time and the flare end time, respectively.} We use these observations to compute the following large-scale, event- and area-integrated quantities: 
\begin{equation} \label{eqphis}
	\Phi_{AR} = \int \limits_\mathrm{AR} |B_{n}| \ dS, \;\;\;\;\;\;\;\;\;\; 
	S_{\mathrm{AR}} = {\int \limits_\mathrm{AR} dS}, \;\;\;\;\;\;\;\;\;\; 
\end{equation}
\begin{equation} \label{eqphis}
	\Phi_\mathrm{ribbon} = \frac{ {\int \limits_\mathrm{I_6} |B_{n} | \ dS} + {\int \limits_\mathrm{I_{10}} |B_{n} | \ dS}}{2}=
	\frac{\Phi_{\rm ribbon}^{(I_6)} + \Phi_{\rm ribbon}^{(I_{10})}}{2},\
\end{equation}
\begin{equation} \label{eqareas}
	S_\mathrm{ribbon} =\frac{ {\int \limits_\mathrm{I_6} \ dS} + {\int \limits_\mathrm{I_{10}} \ dS} }{2}=
	\frac{S_{\rm ribbon}^{(I_6)} + S_{\rm ribbon}^{(I_{10})}}{2},
\end{equation}
\begin{equation} \label{eqratios}
	\overline{B}_\mathrm{AR}=\frac{{\Phi_\mathrm{AR}}}{{S_\mathrm{AR}}},
	\;\;\;\;\;  \overline{B} _\mathrm{ribbon}=\frac{{\Phi_\mathrm{ribbon}}}{{S_\mathrm{ribbon}}},
\end{equation}
\begin{equation} \label{eqratios}
	R_{\Phi} = \frac{\Phi_{\mathrm{ribbon}}}{\Phi_{\mathrm{AR}}} \times100\%,
	\;\;\;\;\; R_{S} = \frac{S_{\mathrm{ribbon}}}{S_{\mathrm{AR}}}\times100\%,
\end{equation}
where $\Phi_{AR}$ and $S_{\mathrm{AR}}$ are the unsigned AR magnetic flux and AR area, 
$\Phi_\mathrm{ribbon}$ and $S_{\mathrm{ribbon}}$ are the unsigned flare-ribbon reconnection flux and flare-ribbon area, $\overline{B}_\mathrm{AR}$ and $\overline{B} _\mathrm{ribbon}$ are the mean AR and ribbon field strengths, and $R_{\Phi}$ and $R_{S}$ are the percentages of the ribbon-to-AR magnetic fluxes and ribbon-to-AR areas, respectively.
 We calculate all the ribbon-related quantities at the flare end time.
The integration $\int \limits_\mathrm{AR} dS$ means summation over the AR region-of-interest and $\int \limits_{I_6} dS $ and $\int \limits_{I_{10}} dS$, summation over the ribbon cumulative pixel mask at $c=6$ and $c=10$ ribbon cutoff thresholds at $t_{\rm final}$. The AR region-of-interest is defined as a $800 \times 800$ pixels ($400 \times 400$ arcseconds) rectangle centered on the AR. The coordinates of the AR center are derived from the HEC mentioned in section~\ref{ssFREL}.


As discussed in Section~\ref{subsec:tempevol}, the uncertainties in the ribbon reconnection flux and the ribbon area are obtained by varying the threshold of minimum ribbon brightness $c$ from $6$ to $10$ times the median background intensity.
The errors in the {\it unsigned} reconnection flux and ribbon area are then 
\begin{align}
\Delta \Phi_\mathrm{ribbon} = \frac{{\Phi^{(I_6)}_{\rm ribbon}} - \Phi^{(I_{10})}_{\rm ribbon}}{2}, \\
\Delta S_\mathrm{ribbon} = \frac{{S^{(I_6)}_{\rm ribbon}} - S^{(I_{10})}_{\rm ribbon}}{2}.
\end{align}

Table~\ref{RDBtable} summarizes the event information included in the {\tt RibbonDB} catalogue that is available online.
\begin{table} 
	\caption{Variables in the {\tt RibbonDB}$^a$ catalogue for each of our \nflares flare ribbon events.} 
	\begin{center}
	\begin{tabular}{l  c  l }
\toprule
 \bf{Variable} 		& \bf{Quantity}		& \bf{Description} \\
 \midrule
 {\tt tstart}			& $t_{\rm start}$		& flare start time [UT] \\
 {\tt tpeak}			& $t_{\rm peak}$		& flare peak time [UT] \\
 {\tt tfinal}			& $t_{\rm final}$		& flare end time [UT] \\
 {\tt ixpeak}		& $I_{\rm X,peak}$		& peak 1-8\AA{} X-ray flux [W~m$^{-2}$] \\
 {\tt lat}			& $lat$				& flare location latitude [deg] \\ 
 {\tt lon}			& $lon$				& flare location longitude [deg] \\
 {\tt arnum}		& $AR_{\rm number}$	& AR number \\
 {\tt phi\_ar}		& $\Phi_{AR}$ 			& total AR unsigned flux [Mx] \\
 {\tt phi\_rbn}		& $\Phi_\mathrm{ribbon}$ & total unsigned rec. flux [Mx] \\
 {\tt dphi\_rbn}	& $\Delta \Phi_{\rm ribbon}$ & recon. flux uncertainty [Mx] \\ 
 {\tt s\_ar}			& $S_{\mathrm{AR}}$ 	& AR area [cm$^2$] \\
 {\tt s\_rbn}		& $S_{\mathrm{ribbon}}$ & ribbon area [cm$^2$] \\
 {\tt ds\_rbn} 	& $\Delta S_{\rm ribbon}$	& ribbon area uncertainty [cm$^2$] \\ 
 {\tt r\_phi}		& $R_{\Phi}$ 			& \% recon. flux to AR flux \\
 {\tt r\_s}		 & $R_{S}$ 			& \% ribbon area to AR area \\
\bottomrule
\end{tabular}
\label{RDBtable} 
\end{center}
\footnotesize{$^a$ \url{http://solarmuri.ssl.berkeley.edu/~kazachenko/RibbonDB/}} 
\end{table}
\subsection{Statistical Analysis}\label{stat}

To quantitatively describe the relationship between different properties of flares and ARs,
e.g. $\mathbb{X}$ and $\mathbb{Y}$, we use Spearman ranking correlation 
coefficient, $r_s(\mathbb{X},\mathbb{Y})$. Unlike the Pearson correlation coefficient that
measures {\it linear} relationship between two variables---and therefore is not suitable for 
non-linearly related variables---the Spearman rank correlation provides a measure of a {\it monotonic} 
relationship between two variables. To estimate the errors in the correlation coefficient due to 
sampling, we use a bootstrap method: for that we repeatedly calculate ${r_{s}}$ for 
$N$-out-of-$N$ randomly selected data pairs with replacement, and then estimate mean and 
standard deviation as $\bar{r}_s \pm \Delta r_s$ (\citet{Wall2012}, Chapter 6). From this point forward, we will refer to the 
mean Spearman correlation $\bar{r}_s$ as simply the correlation coefficient ${r}_s$ and its 
standard deviation $\Delta r_s$ as the correlation's statistical uncertainty.

We describe the qualitative strength of the correlation using 
the following guide for the absolute value of $r_s$: 
$r_s \in [0.2,0.39]$ -- weak, 
$r_s \in [0.4,0.59]$ -- moderate, 
$r_s \in [0.6,0.79]$ -- strong, and 
$r_s \in [0.8,1.0]$ -- very strong. 
When the correlation coefficient is moderate or greater ($r_s(\mathbb{X},\mathbb{Y})>0.4$), we 
fit the relationship between $\mathbb{X}$ and $\mathbb{Y}$ with a power-law function
\begin{equation}
\mathbb{Y} = a \mathbb{X}^b \ .
\end{equation}
We use Levenberg-Marquardt non-linear least-squares minimization method to find scaling factor $a$ and exponent $b$. 
\section{Results}\label{results}	
\subsection{Peak X-Ray Flux vs. Flare Ribbon and AR Properties}\label{results}

\begin{table*}[tbh!]
\caption{Action-region and flare-ribbon properties, $\mathbb{X_\mathrm{AR}}$ and $\mathbb{X_\mathrm{ribbon}}$, where $\mathbb{X}$ is either magnetic flux $\Phi$, area $S$, mean magnetic field $\overline{B}$, ribbon-to-AR fractions of area $R_S$ or magnetic flux $R_\mathrm{\Phi}$. The typical range of each quantity is described as the $20^{th}$ to $80^{th}$ percentile $\mathbb{X}[P_{20},P_{80}]$. The relationship between $\mathbb{X}$ and the peak X-ray flux is characterized by the correlation coefficient $r_s(\mathbb{X},I_\mathrm{X,peak})$; for variables with $r_s>0.4$ we find coefficient $b$ in fit $I_\mathrm{X,peak}=a\mathbb{X}^b$. For more details see the Figures.} 
\begin{center}
\begin{tabular}{ l l l   l   l  c   l}
\toprule
 & \multicolumn{2}{c}{\textbf{ACTIVE~REGIONS}} & \multicolumn{3}{c}{\textbf{FLARE~RIBBONS}} &\\  \cmidrule(r){2-3}  \cmidrule(l){4-6}
Quantity & Typical range & Correlation & Typical range & Correlation & $I_\mathrm{X,peak}$ & Figure \\
 $\mathbb{X}$ & $\mathbb{X}_\mathrm{AR}[P_{20},P_{80}]$ & {$r_s(\mathbb{X}_\mathrm{AR},I_\mathrm{X,peak}) $} & $\mathbb{X}_\mathrm{ribbon}[P_{20},P_{80}]$ & $r_s(\mathbb{X}_\mathrm{ribbon},I_\mathrm{X,peak})$ & $\propto$ & \\
\midrule
 $\Phi$ & $[28,64]\times10^{21}$ Mx & $0.22\pm0.01$ &$[5.4,21]\times10^{21}$ Mx & $0.66\pm0.01$&$ 
 \Phi_\mathrm{ribbon}^{1.53}$ & Fig.~\ref{fig02} \\
$S$ & $[78,183]\times10^{18}$ cm$^2$ & $0.14\pm0.02$ &  $[1.1,3.7]\times10^{18}$ cm$^2$ & $0.68\pm0.01$&$ S_\mathrm{ribbon}^{1.57}$ & Fig.~\ref{fig03} \\
 $\overline{B}$ &$[310,442]$ G & $0.21\pm0.02$  &$[408,675]$ G & $0.24\pm0.02$ & -- & Fig.~\ref{fig04} \\
 $R_S$ &--  & -- &$[0.9,3.4]$ \%& $0.53\pm0.01$  & $ R_\mathrm{\Phi}^{1.7}$  & -- \\
$R_\mathrm{\Phi}$ &--  & -- &$[1.3,5.1]$ \% & $0.54\pm0.01$  & $ R_\mathrm{\Phi}^{1.9}$ & Fig.~\ref{fig05}\\
\bottomrule
%
\end{tabular}
\label{maintab} 
\end{center} 
\end{table*}

\begin{figure*}[htb!]
	\centerline{ \includegraphics[width=0.98\textwidth]{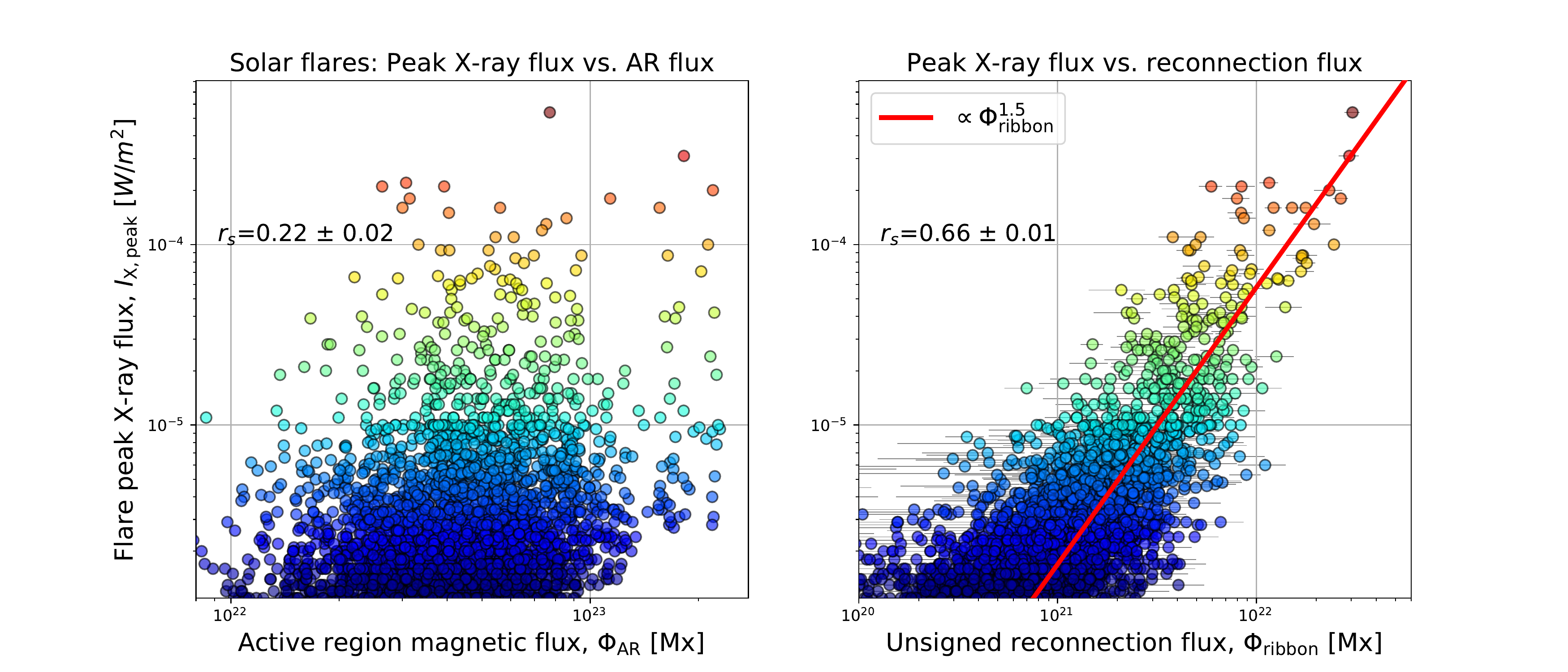} } 
	\caption{{Scatter plots of peak X-ray flux versus unsigned AR 
	magnetic flux and flare reconnection flux. The Spearman correlation 
	coefficients for these cases are listed in each panel. The power-law 
	relationship $I_\mathrm{X,peak} \propto \Phi_\mathrm{ribbon}^{1.5}$ is 
	shown in red. \\}}
 \label{fig02} 
 \end{figure*}

\begin{figure*}[htb!]
	\centerline{ \includegraphics[width=1.0\textwidth]{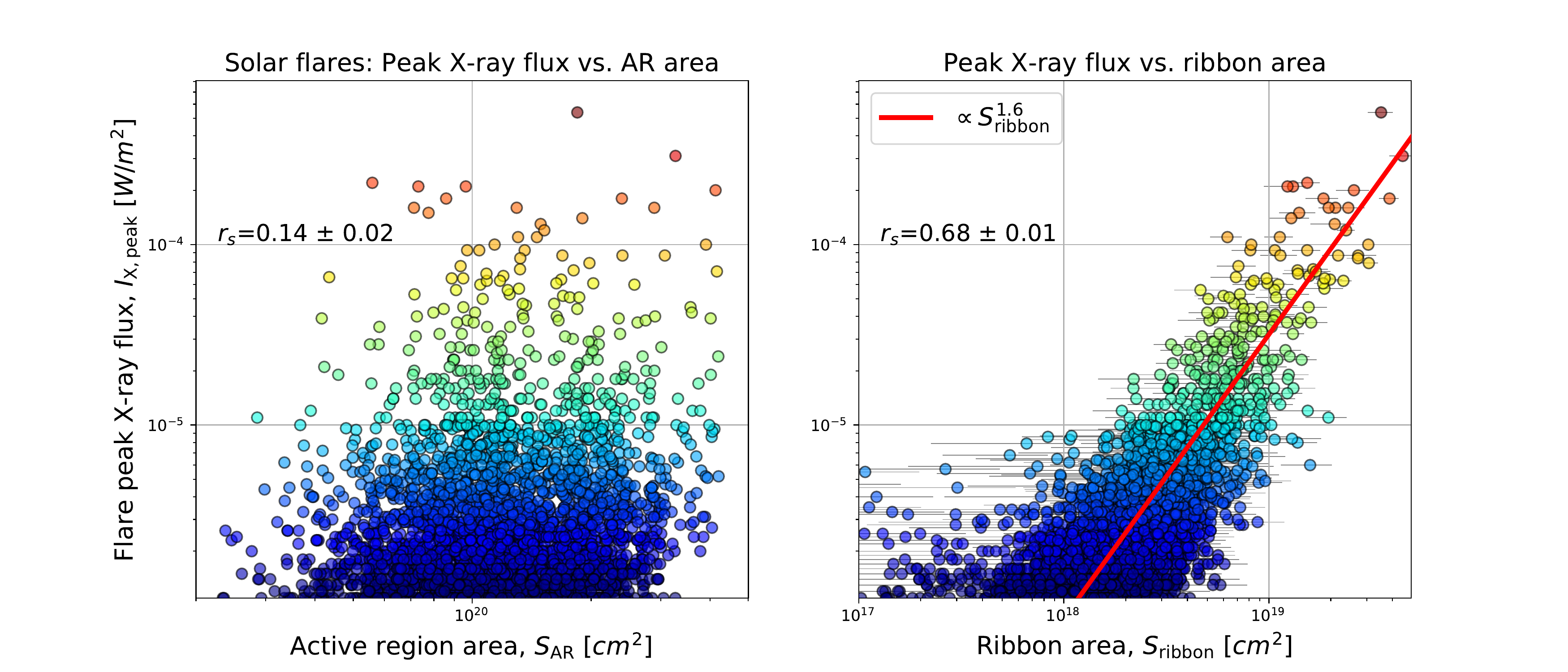} }  
	\caption{{ Scatter plots of peak X-ray flux versus AR area 
	and ribbon area. The Spearman correlation coefficients for these cases 
	are listed in each panel. The power-law relationship is shown for in red. \\}}
	\label{fig03} 
\end{figure*}

\begin{figure}[!b]
	\centerline{ \includegraphics[width=0.50\textwidth]{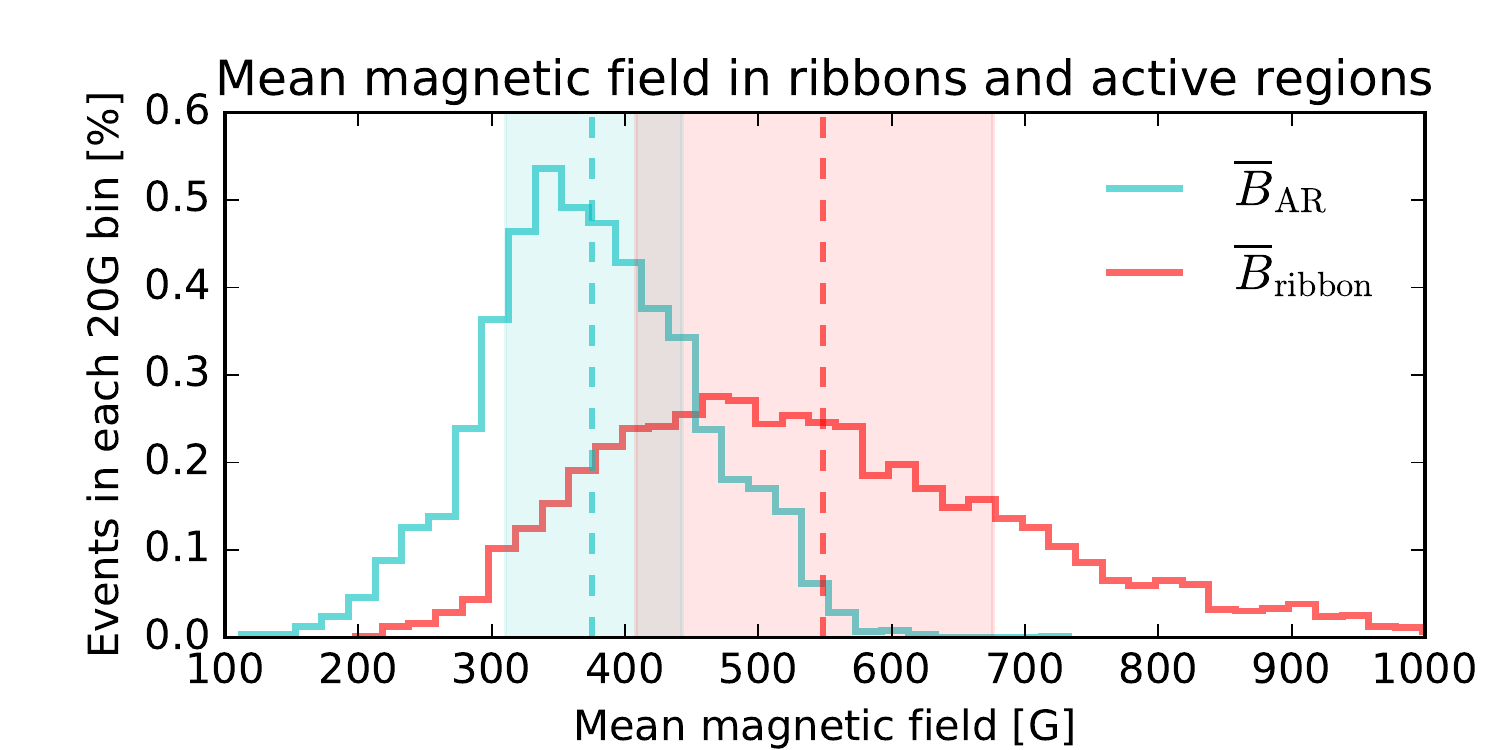} } 
 	\caption{{Histograms for mean magnetic field swept by ribbons (red) and of 
	the whole AR (blue) for all events in dataset. Shaded areas show ranges of magnetic fields 
	where $20\%$ to $80\%$ of events reside. \\}}
 	\label{fig04} 
\end{figure}

\begin{figure*}
\centerline{ \includegraphics[width=1.0\textwidth]{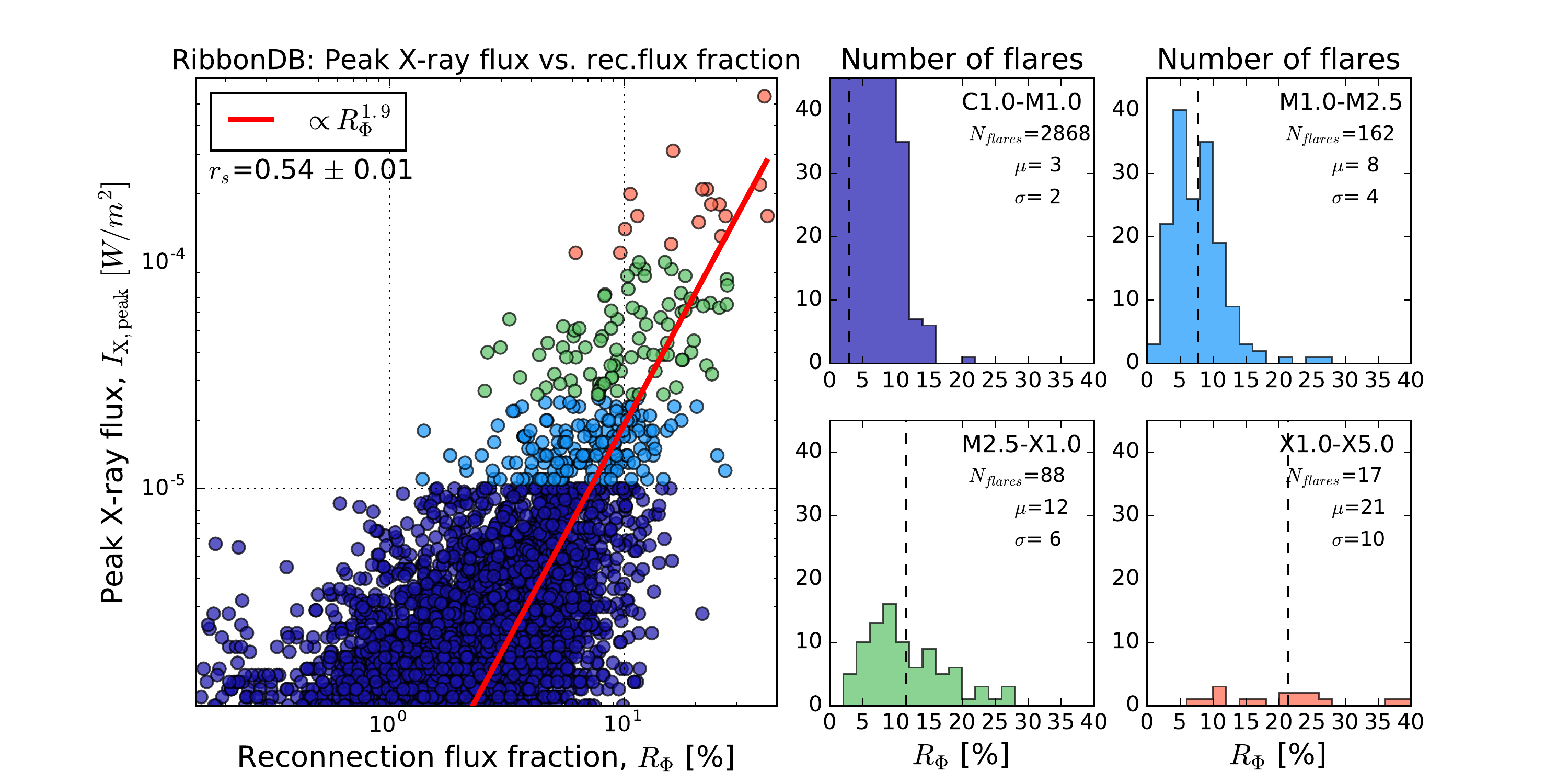} } 
 \caption{{{\it Left panel:} peak X-ray flux versus reconnection flux fraction; {\it four panels on the right:} distribution of number of flares of a certain class versus reconnection flux fraction. $\mu$ and $\sigma$ are mean and the standard deviation of each distribution.\\ }}
 \label{fig05} 
 \end{figure*}

Table~\ref{maintab} summarizes the properties of ARs and flare ribbons listed in Table~\ref{RDBtable}, their range, and correlation coefficient with the GOES peak X-ray flux.
The `Active Regions' column lists the AR unsigned magnetic flux, area, and the mean magnetic field: $\Phi_{AR}$, $S_{AR}$, and $\overline{B}_{AR}$. The `Flare 
Ribbons' column lists the reconnection flux, flare ribbon area, and the mean magnetic field swept by the ribbons: $\Phi_{\rm ribbon}$, $S_{\rm ribbon}$, and 
$\overline{B}_{\rm ribbon}$. 
The bottom row shows the fractions of magnetic flux and area of the whole AR involved in the flare reconnection, $R_{\Phi}$ and $R_{S}$.
We discuss each of these relationships further in text and in Figures~\ref{fig02}--\ref{fig05}.

Figure~\ref{fig02} shows the scatter plots of the flare peak X-ray flux versus the total 
AR unsigned magnetic flux and the flare ribbon 
reconnection flux at $t_{\rm final}$: $I_\mathrm{X,peak}$ vs. 
$\Phi_\mathrm{AR}$, left panel, and $I_\mathrm{X,peak}$ vs. $\Phi_\mathrm{ribbon}$, right 
panel. 
While the flare peak X-ray flux has very little correlation with the AR magnetic 
flux, $r_s(I_\mathrm{X,peak},\Phi_\mathrm{AR})=0.22\pm0.01$, it is strongly correlated with the flare ribbon reconnection 
flux, $r_s(I_\mathrm{X,peak},\Phi_\mathrm{ribbon})=0.66\pm0.01$. The correlation is strong: 
$r_s=0.66\pm0.01$. 
The power law fit to $I_\mathrm{X,peak}$ vs. $\Phi_\mathrm{ribbon}$ yields 
$I_\mathrm{X,peak} \propto \Phi_\mathrm{ribbon}^{1.5}$.  

If we restrict our analysis to stronger flares, M1-class and above, we find a weaker correlation coefficient with larger standard deviation: $r_s(I_\mathrm{X,peak},\Phi_\mathrm{ribbon},  >M1.0)=0.51\pm0.05$. This weakening is a result of the well-known in statistics problem of {\it range restriction} \citep{Pearson1903}, rather than a consequence of different physical processes governing smaller and larger flares. This effect reduces  correlation coefficients for flares of smaller range of flare classes, e.g. flares larger than M1 or flares smaller than C5. For the same reason we would expect larger correlation between the flare peak X-ray flux and the reconnection flux if we include flare classes beyond the \verb+RibbondDB+ range.


%
Additionally, we investigated the difference between using the normal component of the magnetic $B_n$ field derived from the line-of-sight (LOS) versus the full vector magnetograms (as we do here).
The normal component is then derived as $B_\mathrm{n}=B_\mathrm{LOS}/\cos{\theta_j}$, where $\theta_j$ is the angular distance between the central meridian and the pixel $(x_i, y_j)$. 
We find that the relationships between $I_\mathrm{X,peak}$ and $\Phi_\mathrm{AR}$ and $\Phi_\mathrm{ribbon}$, their correlation coefficients, and the power-law exponents using $B_\mathrm{LOS}$ are within the uncertainties of the estimates using the vector magnetic fields shown above.
 
Figure~\ref{fig03} shows the scatter plots of flare peak X-ray flux versus total 
AR area ($I_\mathrm{X,peak}$ vs. $S_\mathrm{AR}$, left panel) and 
cumulative flare ribbon area at $t_{\rm final}$ ($I_\mathrm{X,peak}$ 
vs. $S_\mathrm{ribbon}$, right panel). 
Here, we find an even weaker correlation between the peak X-ray flux and the AR area 
($r_s=0.14\pm0.02$) and slightly stronger correlation between the peak X-ray flux 
and the cumulative flare ribbon area ($r_s=0.68\pm0.01$) than the AR flux and the ribbon 
reconnection flux, respectively.

We further examine whether the mean magnitude of the normal magnetic field 
swept by the flare ribbons, $\overline{B} _\mathrm{ribbon}$, is substantially different than the mean field of the 
whole AR, $\overline{B}_\mathrm{AR}$. %
Neither $\overline{B}_\mathrm{AR}$ nor $\overline{B} _\mathrm{ribbon}$ show 
anything more than a weak correlation with flare peak X-ray flux ($0.21\pm0.02$ and $0.24\pm0.02$, respectively).
This does not contradict though the known association between strong gradients in the normal magnetic field across the AR polarity inversion line and the AR's flare and CME productivity \citep[e.g.][and references therein]{Welsch2008b}. 
Figure~\ref{fig04} plots the distributions of $\overline{B}_{AR}$ and 
$\overline{B}_{\rm ribbon}$ for \nflares ribbon events. We find that the average of the $\overline{B}_{\rm ribbon}$ field 
strength distribution is $100$ to $200$~G higher than the average of the 
$\overline{B}_{AR}$ distribution. 
The range between the $20$--$80$ percentiles 
for $\overline{B}_{AR}$ is $310-442$~G (light blue shaded region), whereas the 
corresponding percentile range for $\overline{B}_{\rm ribbon}$ is $408-675$~G 
(light red shaded region).
The Figure~\ref{fig04} results confirm, in a statistical sense, that the magnetic fields 
that participate in the flare reconnection tend to be stronger fields than the AR as a whole. 
 
Lastly, in Figure~\ref{fig05}, we examine the relationship between the fraction of the AR magnetic flux that 
participates in the flare reconnection and the flare peak X-ray flux: $I_\mathrm{X,peak}$ versus $R_\mathrm{\Phi}$ (see Equation~\ref{eqratios}). 
The left panel of Figure~\ref{fig05}  shows that the flare peak X-ray flux exhibits a moderate correlation with $R_\mathrm{\Phi}$ ($r_s=0.54\pm0.01$). The fit to the power-law relationship between the two quantities yields $I_\mathrm{X,peak} \propto R_\mathrm{\Phi}^{1.9}$. 
The right panels of Figure~\ref{fig05} show the histogram distributions 
of $R_\mathrm{\Phi}$ in four ranges of flare-class: C1--M1 (dark blue), M1--M2.5 (light 
blue), M2.5-X1 (green), and X1-X5 (red). The distribution mean (vertical 
dashed line) and standard deviation are listed in each panel with the 
total number of events in the class-range. For C1--M1 flares, the mean and standard deviation of $R_\mathrm{\Phi}$ is $3\pm2\%$; 
for M1--M2.5 flares, $8\pm4\%$; for M2.5-X1 flares, $12\pm6\%$; and 
for X-class flares, $21\pm10\%$. To summarize, both the mean and the width of the distributions of the reconnected flux fraction increase with the flare class strength. 

 \subsection{Occurrence Frequencies of Peak X-ray Flux, Reconnection Flux, and Flare Magnetic Energy}\label{results}

To describe the statistical properties of solar and stellar flares, many authors have looked at the frequency distributions of various flare and CME parameters, such as flare duration, peak hard X-ray and soft X-ray fluxes, total magnetic energy and its thermal, radiative, and the electron contributions \citep{Drake2013,Maehara2015,Harra2016,Notsu2016}. Since all of these quantities are products of the flare reconnection process \citep{Forbes2000}, it is imperative to understand the quantitative distribution of flare reconnection fluxes and their associated magnetic energy release. 

Following \citet{Aulanier2013}, we estimate the energy released in the flare in terms of the properties of the reconnected magnetic field as
\begin{equation}
	E_\mathrm{flare} \sim
	f E_\mathrm{mag} \sim
	f \frac{\overline{B}_\mathrm{ribbon}^2}{8\pi} V \sim
	f \frac{\overline{B}_\mathrm{ribbon}^2}{8\pi} \frac{S_\mathrm{ribbon}}{2}^{3/2},\
\label{flareeneq} \end{equation}
where $f$ is the fraction of magnetic energy released as flare energy and $V$ is the volume of reconnecting magnetic fields expressed in terms of the flare ribbon area $V\sim(S_\mathrm{ribbon}/2)^{3/2}$. 
We divide $S_\mathrm{ribbon}$ by a factor of $2$ to derive signed from unsigned quantities. We also assume that all the non-potential magnetic energy is released by the flare, i.e. $f=1$. We note that Equation~\ref{flareeneq} differs from flare energy estimates of \citet{Shibata2013} and \citet{Maehara2015} for stellar flares. Here, instead of the properties of the whole AR, we only use the flaring portion of the AR, as defined by the flare ribbons.

 \begin{figure*}[htb]
 \begin{tabular}{c} 
	\centering \includegraphics[width=1.0\textwidth]{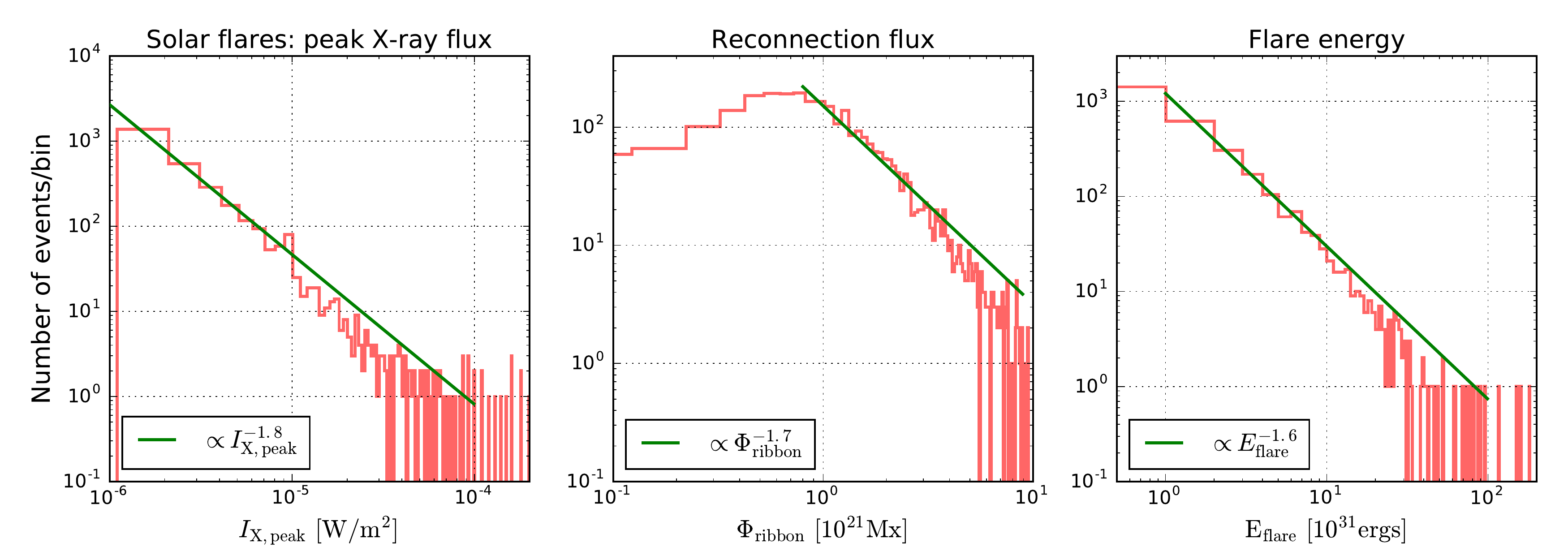}  
	\end{tabular}
	\figcaption{Histograms of the GOES peak X-ray flux, ribbon unsigned reconnection flux, 
	and an estimate of the flare magnetic energy. 
 	\label{fig12}}
\end{figure*}

Figure~\ref{fig12} shows the occurrence frequencies for the GOES peak X-ray flux, ribbon unsigned reconnection flux, and the flare magnetic energy proxy: $I_\mathrm{X,peak}$, $\Phi_\mathrm{ribbon}$ and $E_\mathrm{flare}$.
The number of solar flares, proportional to their occurrence frequency, decreases dramatically as the flare energy increases. 
The fit to the power-law dependence of the peak X-ray flux distribution yields $dN/dI_{X,peak} \propto I_{X,peak}^{-1.8}$  for $I_{X,peak} \in [10^{-6},5\times10^{-4}]~\mathrm{W~m^{-2}}$. The fit to the power-law dependence of the reconnection flux distribution yields $dN/d\Phi_{\rm ribbon} \propto \Phi_{\rm ribbon}^{-1.7}$ for $\Phi_{\rm ribbon}\in [8\times 10^{20},10^{22}]$~Mx. 
The fit to the power-law dependence of the distribution of flare energy, $E_\mathrm{flare}$, yields $dN/dE_\mathrm{flare} \propto E_\mathrm{flare}^{-1.6}$ for  $E_\mathrm{flare}\in[10^{31},10^{33}]$~erg. 

We also compare the distribution of our flare energy proxy derived from the reconnection flux with previous results for both solar and stellar flares.
For solar flares, \citet{Crosby1993}, \citet{Shimizu1995}, \citet{Aschwanden2000}, and \citet{Qiu2004} have examined the occurrence frequency as a function of flare energy over $E_\mathrm{flare} \sim 10^{24-32}$~erg which includes micro-flares and nano-flares. These distributions, obtained from EUV and soft and hard X-ray emission, are generally well-described by a simple power-law function, $dN/dE_\mathrm{flare} \propto E_\mathrm{flare}^{-\alpha}$, with an index, $\alpha$, of 
$1.5-1.9$. 
For much larger stellar flares, $E_\mathrm{flare}\sim10^{32-36}$ erg, \citet{Maehara2015} and \citet{Notsu2016} found a power-law dependence, $\alpha=1.5\pm0.1$. In these estimates, the area of the starspot group was used as a proxy of stored magnetic energy, $E_\mathrm{flare}$. 
Combining the solar and stellar flare results, \citet{Shibata2013} suggested that the frequency distribution should follow a universal power law index, $\alpha=1.8$, for $E \sim 10^{24-36}$~erg. 

In this paper, we use high spatial resolution observations of flares on the Sun to estimate only the magnetic energy released during the flare, not the total AR magnetic energy. As shown in Figure~\ref{fig02}, the total AR flux or area, is only weakly correlated with the flare class. Hence, the total AR flux and area are, physically, not the best proxy for $E_\mathrm{flare}$. The exponent for our distribution of flare energies is $\alpha=1.6$, well within the range of exponents previously found for both stellar and solar flares $(1.5-1.9)$.

\section{Discussion}\label{disc} 
We have shown, in Figures~\ref{fig02}--\ref{fig05}, that while the GOES peak X-ray flux is only weakly correlated with the AR quantities ($\Phi_{AR}$, $S_{AR}$), it is strongly correlated with the equivalent quantities derived from the flare ribbon observations ($\Phi_{\rm ribbon}$, $S_{\rm ribbon}$, $R_{\Phi}$). 

Previous studies, e.g. \citet{Barnes2008}, have established that the more unsigned flux there is within an AR, the higher its overall rate of flare production, and the more likely the AR is to produce large (e.g., X-class) flares. 
%
%
Given the low correlation between the AR flux and the flare class that we find, how can the greater likelihood of large flares from large ARs be understood? 
If the magnetic reconnection processes responsible for the flare are universal (i.e. largely insensitive to action region size), then one may expect a universal distribution of flare frequencies as a function of energy \citep[e.g.,][]{Wheatland2000,Wheatland2010}. In this way, a larger AR is more likely to produce flares of all classes and the likelihood of large flares will be enhanced relative to smaller ARs. 

Understanding the stronger correlation between reconnection flux and peak X-ray flux is fairly intuitive. The peak X-ray intensity is a measure 
of the $\sim$1--10MK temperature emission response of the solar atmosphere to the rapid energy deposition supplied by magnetic fields that reconnected. It makes sense that the physical properties more closely associated with the flare reconnection process---our ribbon quantities---are more correlated with the resulting X-ray intensities than properties of the whole AR derived from the normal component of the AR magnetic fields.

\citet{Warren2004} analyzed the hydrodynamic response of impulsively 
heated flare loops and found that the peak soft X-ray flux scales approximately with the flare energy as 
$I_\mathrm{X,peak} \propto E_\mathrm{flare}^{1.75} \ V^{-0.75} \ L^{-0.25}$,
where $V$ is the flare volume and $L$ is the flare loop length. 
Writing $V$ and $L$ in terms of the flare ribbon area as $V\sim L^3\sim(S_\mathrm{ribbon}/2)^{3/2}$ and using the Equation~\ref{flareeneq} estimate for $E_\mathrm{flare}$, we can rewrite the above relation as a function of our flare ribbon quantities:
\begin{equation}
I_\mathrm{X,peak} \propto \frac{ E_\mathrm{flare}^{1.75} }{ S_{\rm ribbon}^{1.25} } 
     \sim \frac{ \Phi_{\rm ribbon}^{3.5} }{ S_{\rm ribbon}^{2.125} }. \ 
\end{equation}
Expressing the ribbon area in terms of the reconnection flux as $\Phi_{\rm ribbon} = \overline{B}_{\rm ribbon} S_{\rm ribbon}$, we derive a theoretical scaling for the peak X-ray flux,  
\begin{equation}
I_\mathrm{X,peak} \propto \Phi_{\rm ribbon}^{1.375}. 
\end{equation}
This estimate is remarkably consistent with our observed 
($I_\mathrm{X,peak}$, $\Phi_{\rm ribbon}$) relationship despite the 
approximations made above about the flare loop geometry.
The power-law fit is $I_\mathrm{X,peak} \propto \Phi_{\rm ribbon}^{1.5}$ (see Figure~\ref{fig02}). 

We also speculate that due to the Neupert effect \citep{Neupert1968,Veronig2002}, the derived correlation between the reconnection flux and the peak soft X-ray flux will lead to a strong correlation between the peak reconnection flux rate and the peak hard X-ray flux, in agreement with earlier studies of individual events \citep[e.g.][]{Qiu2002,Veronig2015}.

\section{Conclusions}\label{conc}
Since solar flares release energy stored in the magnetic field, the main property that describes the flare process is the amount of magnetic flux that reconnects, i.e., the reconnection flux. 
Previous estimates of the reconnection fluxes from observations of flare ribbon evolution were performed for only a limited number of events. 
The launch of \emph{SDO}, with the HMI and AIA instruments onboard, provided the first opportunity to compile a much larger sample of flare-ribbon events. Taking advantage of this newly available data and our ribbon analysis techniques, we assembled a new {\tt RibbonDB} catalogue of \nflares events. The {\tt RibbonDB} catalogue contains flare ribbon and AR properties (Table~\ref{RDBtable}) for every flare of GOES class C1.0 and greater within 45 degrees of the central meridian, from April 2010 until April 2016. 

We analyze the properties of the ARs and flare ribbons in each event, including the AR magnetic flux, AR area, and the mean AR field strength, and the flare reconnection flux, flare ribbon area, and the mean strength of the fields swept by ribbons. We compare these quantities with the GOES peak X-ray flux as a proxy of radiative flare energy. Our findings are as follows.

\begin{enumerate}
\item We find strong statistical correlations between the flare peak X-ray flux and our derived flare ribbon quantities, cumulative ribbon area and reconnection flux: Spearman correlation coefficient $r_s(I_\mathrm{X,peak},S_\mathrm{ribbon})=0.68\pm0.01$ and $r_s(I_\mathrm{X,peak},\Phi_\mathrm{ribbon})=0.66\pm0.01$,  respectively. In contrast, the correlation between the peak X-ray flux and the corresponding AR quantities is weak: $r_s(I_\mathrm{X,peak},S_\mathrm{AR})=0.14\pm0.02$ and $r_s(I_\mathrm{X,peak},\Phi_\mathrm{AR})=0.22\pm0.01$. 

\item We find the power-law relationship between the peak X-ray flux and the ribbon reconnection 
flux to be  $I_\mathrm{X,peak} \propto \Phi_\mathrm{ribbon}^{1.5}$. 
This exponent value are consistent with the \citet{Warren2004} scaling law derived 
from hydrodynamic simulations of impulsively heated flare loops, thus indicating that 
the energy released during the flare as soft X-ray radiation originates from the free 
magnetic energy stored in the magnetic field released during reconnection.

\item We find a moderate correlation between the flare peak X-ray flux and the percentage of magnetic flux that gets reconnected: $r_s(I_\mathrm{X,peak},R_\mathrm{\Phi})=0.54\pm0.01$. Both the mean and the width of the distributions of the reconnected flux fraction increase with the flare class strength.

\item We find that the occurrence frequencies of the flare peak X-ray fluxes, reconnection fluxes, and the flare energies can be fit with the same power law, $dN/dX\propto X^{-\alpha}$, with a power law index, $\alpha\in[1.6,1.8]$. These results are consistent with previous studies of solar and stellar flares derived from AR properties. 
 \end{enumerate}
  
This study is the first large-sample statistical analysis of the flare reconnection fluxes and their relationship with other flare and AR properties. While here, we focus on the cumulative reconnection properties, in the second paper we plan to extend our analysis to the statistical properties of the temporal evolution of flare ribbons, such as the ribbon speed and the reconnection flux rate. We believe that such a statistical approach is very beneficial since it enables us to investigate general trends that may be overlooked in case studies of individual events. 

The RibbonDB catalogue is available \href{http://solarmuri.ssl.berkeley.edu/~kazachenko/RibbonDB/}{online} in a CSV and an IDL SAV file formats, and can be used for a wide spectrum of quantitative studies in the future. For example, comparison of reconnection fluxes with HXR emission, SEP fluxes, CME and magnetic cloud properties would be valuable to clarify the relationship between the flares and ICMEs/CMEs,  e.g. extending the \citet{Gopalswamy2017} analysis to a much larger number of events. Analysis of the outliers in the derived trends, for example, events with large X-ray flux but small reconnection flux and vice versa, would be very interesting. 

\acknowledgments
We thank Marc DeRosa and the AIA team for providing us with the SDO/AIA data. We thank the HMI team for providing us with the vector magnetic field SDO/HMI data. We thank George Fisher for reading and correcting the manuscript. We are grateful to Jiong Qiu and Dana W. Longcope for helpful discussions. We thank US taxpayers for providing the funding that made this research possible. We acknowledge support from NASA H-GI ODDE NNX15AN68G (M.D.K., B.T.W, B.J.L.), National Science Foundation, SHINE, AGS 1622495 (M.D.K., B.J.L.), NASA award NAS5-02139 (HMI, X.S.), Coronal Global Evolutionary Model (CGEM) award NSF AGS 1321474 (M.D.K., B.T.W., B.J.L.) and CGEM NASA award NNX13AK39G (X.S.).

\end{document}